\pdfoutput=1
\documentclass[showpacs,amsmath,amssymb,aps,longbibliography,twocolumn,pra]{revtex4-2}%
\usepackage{graphicx}

\usepackage[utf8]{inputenc}

\usepackage{graphicx}

\usepackage{etoolbox}
\makeatletter
\pretocmd{\NAT@open}{\begingroup\color{\@citecolor}}{}{}
\apptocmd{\NAT@close}{\endgroup}{}{}
\makeatother

  {\acknowledgments}%
  {\endacknowledgments}

\usepackage{xr}
\usepackage{braket}
\usepackage{bbm}

\usepackage{placeins}
\usepackage{soul}
\usepackage[usenames]{color}
\definecolor{dark-blue}{rgb}{0,0.2,0.6}

\usepackage[english]{babel}
\usepackage{enumerate}
\usepackage{pifont}
\usepackage{amsfonts}
\usepackage{siunitx}
\usepackage{gensymb}

\usepackage{amssymb}
\usepackage{wasysym}
\usepackage{physics}
\usepackage{bbold}

\usepackage{amsmath}
\usepackage{mathtools}

\usepackage{upgreek}

\usepackage{xspace}
\usepackage{titletoc}

\usepackage[colorlinks,%
            pdfusetitle,%
            urlcolor=dark-blue,%
            citecolor=dark-blue,%
            linkcolor=dark-blue]{hyperref}

\newcommand{\be}{\begin{equation}}
\newcommand{\ee}{\end{equation}}

\newcommand{\contentsitem}[2]{%
  \item[\ref{#1}]%
  {\rightskip=2em plus 1fil \parfillskip=-\rightskip
   \hyperref[#1]{#2}\nobreak\hfill\nobreak
   \makebox[2em][r]{\pageref{#1}}\par}}

\newcommand{\asciimathunit}[1]{\ensuremath{\,\text{#1}}}

\newcommand{\ms}{\asciimathunit{ms}}

\newcommand{\subfigref}[2]{\hyperref[fig:#1]{\ref*{fig:#1}(#2)}}

\renewcommand{\eqref}[1]{Eq.\,\ref{#1}}

\usepackage{enumitem}

\usepackage{dblfnote}
\usepackage{lipsum}

\DFNalwaysdouble

\usepackage{xfp}

\newcommand{\specklecal}{14.22}

\begin{document}
\title{Exploring anomalous Floquet-Anderson insulators \\ with ultracold atoms in optical lattices}

\author{Johannes~Arceri}
\email{johannes.arceri@physik.uni-muenchen.de}
\author{Moritz~Hornung}
\author{Dizhou~Xie}
\author{Christoph~Braun}
\author{Monika~Aidelsburger}
\email{monika.aidelsburger@mpq.mpg.de}
\affiliation{Max-Planck-Institut f\"{u}r Quantenoptik, 85748 Garching, Germany}
\affiliation{Munich Center for Quantum Science and Technology (MCQST), 80799 M\"{u}nchen, Germany}
\affiliation{Fakult\"{a}t f\"{u}r Physik, Ludwig-Maximilians-Universit\"{a}t, 80799 M\"{u}nchen, Germany}

\date{\today}

\begin{abstract}
In two-dimensional static systems, a key signature of topological robustness is quantized edge transport in the presence of disorder. Crucially, the latter relies on the existence of at least one delocalized bulk state. Therefore, disorder-driven Anderson localization of an entire band and a non-trivial Chern number are mutually exclusive. Beyond the static paradigm of the bulk-boundary correspondence, periodic Hamiltonian engineering grants access to anomalous Floquet topological insulators, which host topologically protected edge states despite a seemingly trivial bulk. In such a scenario, the coexistence of chiral edge modes and Anderson localization has been predicted in a genuinely out-of-equilibrium phase known as the anomalous Floquet-Anderson insulator. Despite considerable theoretical efforts, no experimental realization has been reported to date on any platform. Here, we leverage real-space detection of atomic wave packet dynamics in the bulk of driven optical honeycomb lattices and in proximity to a topological interface to reveal signatures of an anomalous Floquet-Anderson insulating phase. In the anomalous Floquet regime, we observe a non-monotonic dependence of bulk expansion on the disorder strength, with a minimum at intermediate disorder that coexists with chiral edge transport and is absent in conventional Haldane systems realized in the same lattice. Scanning the drive parameters across this regime, we find that the contrast of this non-monotonic behavior is correlated with the spectral properties of the clean Floquet band structure. Our results lay the groundwork for further experimental investigations of anomalous Floquet-Anderson insulators, including future realizations in the strongly correlated regime.
\end{abstract}

\maketitle

Geometric phases of electronic wavefunctions have dramatically impacted our understanding of condensed matter systems~\cite{Berry_1984, Simon1983}, leading to the discovery of topological insulators~\cite{TKNN1982, Haldane1988, KaneMele2005, Hasan2010}. Defying Landau's paradigm of local order parameters~\cite{Landau1965}, a topological phase is classified in terms of global invariants, the canonical example being the Chern number $\mathcal{C}$ in two-dimensional systems~\cite{TKNN1982, Hasan2010}. Remarkably, despite characterizing a system's bulk, the Chern number is linked to the existence of chiral modes propagating ballistically along the system's boundary, by virtue of the bulk-boundary correspondence~\cite{Hatsugai1993}. Moreover, disorder does not destroy this physics but enriches it, as exemplified by the stability of integer~\cite{vonKlitzing1986} and fractional~\cite{Stormer1999} quantum Hall plateaus, by the topological Anderson insulator, in which disorder drives a trivial system into a topological regime~\cite{Li2009, Meier_2018, Stutzer_2018, Nakajima_2021}, and by disorder-driven transitions between distinct topological regimes~\cite{Zheng2024, hesse2025probingdisorderdriventopologicalphase}. On the other hand, Anderson localization, the interference-induced suppression of wave transport in a disordered medium~\cite{Anderson_1958, Evers2008}, is fundamentally incompatible with conventional topology in a static system: the Wannier obstruction~\cite{Halperin_1982,Thouless_1984} requires a bulk band to include at least one delocalized state for its Chern number to be non-zero, so that a Chern band cannot localize completely unless its topology is destroyed.

Floquet engineering~\cite{Goldman2014, Bukov_2015, Eckardt2017, Oka_2019, Weitenberg_2021}, i.e., time-periodic modulation of a Hamiltonian $\hat{H}(t)=\hat{H}(t+T)$, has been shown to grant access to so-called anomalous Floquet topological insulators, which admit topologically protected edge states despite vanishing Chern numbers~\cite{Kitagawa_2010, Rudner2013, Nathan2019}. They have been realized with ultracold atoms~\cite{Wintersperger_2020, Zhang2023,asteria2026adiabaticrealizationanomalousfloquet}, photonic waveguide arrays~\cite{Maczewsky_2017}, and acoustic metamaterial ring lattices~\cite{Peng_2016}. Notably, the coexistence of an Anderson-localized bulk with topologically protected edge modes is possible in such a scenario, as the latter are no longer tied to non-vanishing Chern numbers. Hence, periodically driven systems with spatial disorder have been predicted to host an exotic phase known as the anomalous Floquet-Anderson insulator (AFAI)~\cite{Titum_2016, Kundu2020}, in which all bulk states are Anderson-localized while topologically robust chiral edge transport persists. AFAIs support quantized charge pumping in a nonadiabatic framework~\cite{Titum_2016}, hence in the absence of the main assumption underlying conventional Thouless pumps~\cite{Citro_2023}. Furthermore, AFAIs have been speculated to stabilize topological transport alongside many-body localization (MBL) in the strongly correlated regime~\cite{Nathan2019}. Despite these intriguing properties, the experimental realization of AFAIs has so far remained elusive. Observing the phase requires combining a Floquet drive with controlled disorder and probing bulk localization and edge transport in the same system.

Ultracold atoms offer all necessary ingredients. Anderson localization has been extensively investigated on this platform across different dimensionalities~\cite{Billy_2008, Roati_2008, Kondov_2011, Jendrzejewski_2012, Semeghini_2015, White_2020}, relying on imaging of density profiles, tailored transport protocols and time-of-flight detection of momentum distributions, while Floquet-engineered optical lattices host anomalous phases whose edge modes can be probed in real space~\cite{Braun_2024,hesse2025probingdisorderdriventopologicalphase}. In this work, we combine real-space detection of topological edge modes with imaging of wave packet dynamics in the bulk of a driven honeycomb lattice to report experimental signatures of an AFAI phase. From the bulk expansion images, we extract a participation length that characterizes the spatial extent of the atomic cloud and, as our numerical simulations show, follows the participation ratio of the bulk Floquet states. Moreover, the degree of microscopic control over system parameters allows us to directly compare the effect of disorder on anomalous Floquet and conventional Haldane phases in the same lattice.

\begin{figure}[t!]
\centering
\includegraphics{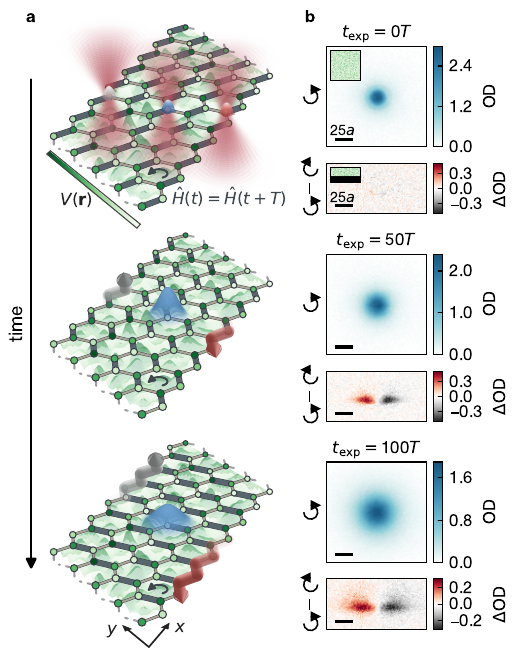}
\caption{\textbf{Wave packet dynamics in the anomalous Floquet regime with spatial disorder.} 
 \textbf{a} Schematic representation of initially tightly confined wave packets and subsequent propagation after release in a periodically modulated honeycomb lattice subject to spatial disorder $V(\mathbf{r})$.
 The modulation scheme results in preferred tunneling along selected lattice directions, clockwise or counter-clockwise around hexagonal plaquettes (circular arrows). Wave packets released from the tweezer in the interior of the lattice (blue) populate bulk Floquet states, whose propagation is hindered by destructive interference between scattering channels. Wave packets released near the boundary (red, grey; separate experiments) populate delocalized edge modes, which propagate chirally along the system boundary despite the disorder potential, with opposite group velocities at opposite edges.
 \textbf{b} Time evolution of the atomic cloud's optical density (OD) as a function of the expansion time $t_\mathrm{exp}$ after release into the disordered ($\overline{V}_D/h=\SI{\fpeval{round(\specklecal * 0.15, 2)}}{\kilo\hertz}$) modulated lattice with frequency $\omega/(2\pi)= \SI{8}{\kilo \hertz}$ and amplitude $m=0.34$, in the absence (top panels) and in proximity (bottom panels) of a repulsive potential step realizing the topological interface. The insets for $t_\mathrm{exp}=0T$ display the combined potential landscape.
 The chiral signal $\Delta$OD is obtained by subtracting the disorder-averaged absorption image for $\kappa=+1$ from that for $\kappa=-1$ (circular arrows). Absorption images are averaged over 100, 200 and 400 disorder realizations for $t_\mathrm{exp}=0T$, $t_\mathrm{exp}=50T$ and $t_\mathrm{exp}=100T$, respectively.}
\label{fig:Fig1}
\end{figure}

\begin{figure*}[t]
\centering
\includegraphics{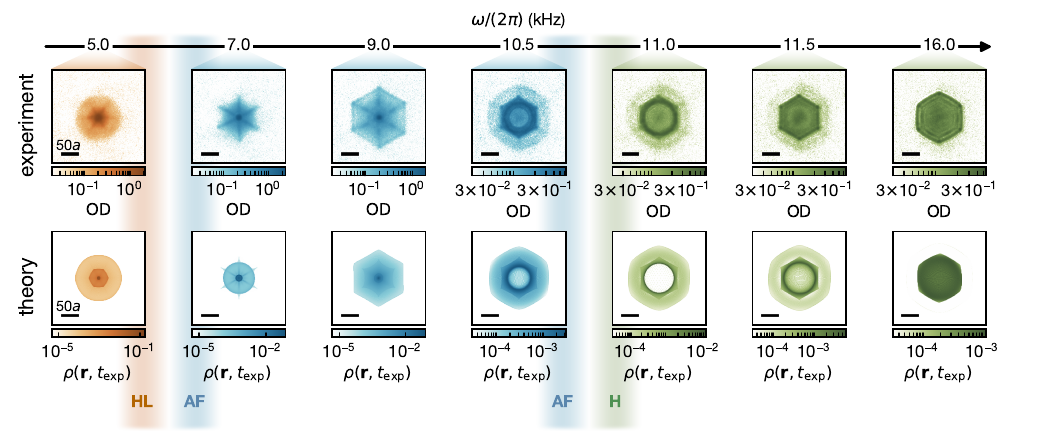}
\caption{\textbf{Expansion dynamics of bulk Floquet states in a tunneling-modulated honeycomb lattice.}
Top row: absorption images of the real-space density distribution in the absence of disorder, $t_\mathrm{exp}=\SI{5}{\milli\second}$ after release from the optical tweezer into the modulated lattice, averaged over 130 experimental repetitions. The modulation amplitude is fixed to $m=0.25$ and the modulation frequency $\omega/(2\pi)$ increases from left to right (arrow). Vertical shadings mark the two topological phase transitions crossed along the cut, from the Haldane-like (HL, orange) regime with $(W_0,W_\pi)=(0,1)$ to the anomalous Floquet (AF, blue) regime with $(W_0,W_\pi)=(1,1)$, and from the AF regime to the Haldane (H, green) regime with $(W_0,W_\pi)=(1,0)$. Bottom row: simulated density distributions $\rho(\boldsymbol{r},t_\mathrm{exp})$ obtained from the group velocities of the two $s$-like Floquet bands resulting from a six-band calculation, weighted by their squared overlap with the initial wave packet (see text and SM). Both rows use logarithmic color scales with ranges adapted to each panel.
}
\label{fig:Fig2}
\end{figure*}

All experiments discussed below investigate the temporal dynamics of tightly confined $^{39}$K Bose-Einstein condensates (BECs) in a periodically driven optical honeycomb lattice. The static lattice potential is realized by interfering three blue-detuned, $s$-polarized laser beams with wavelength $\lambda_L = \SI{745}{\nano\meter}$ at a relative angle of $120^{\circ}$ in the atomic plane, resulting in a lattice spacing of $a=2/({3\sqrt{3}})\lambda_L \approx \SI{287}{\nano\meter}$. The lattice depth is $5.9E_R$, with the recoil energy $E_R=h^2/(2\lambda_L^2m_\mathrm{K}) \approx h \times \SI{9.23}{\kilo \hertz}$, Planck's constant $h$ and the mass of $^{39}$K $m_\mathrm{K}$. The corresponding bare tunneling rate is $J_0=\SI{1.1 }{\kilo \hertz}$. The driving protocol consists of a phase-shifted sinusoidal modulation of the relative intensity of the three lattice beams: $I_j(t)=I_0\left(1-m+m\cos{\left(\omega t-j\kappa(2\pi)/{3}\right)}\right)$, $j\in \{ 1,2,3\}$, where $\omega=2\pi/T$, $m$ and $\kappa= \pm1$ (counter- and clockwise arrows in Fig.~\ref{fig:Fig1}) denote the frequency, amplitude, and chirality of the intensity modulation, respectively~\cite{Wintersperger_2020, Braun_2024, hesse2025probingdisorderdriventopologicalphase}. As a consequence, tunneling amplitudes are sequentially enhanced along different lattice directions, producing a clockwise ($\kappa=-1$) or counter-clockwise ($\kappa=+1$) pattern of unequal couplings around hexagonal plaquettes over each modulation period, in a continuous version of the toy model introduced by Kitagawa et al.~\cite{Kitagawa_2010}. The parameter space spanned by $\omega/(2\pi)$ and $m$ supports several topological phases, including conventional Chern insulators as well as genuinely non-equilibrium anomalous Floquet phases~\cite{Wintersperger_2020, Braun_2024}. 
\\
Disorder is introduced by an optical speckle pattern ($\lambda_{\mathrm{sp}}=\SI{532}{\nano \meter}$), which realizes a static disorder potential $V(\mathbf{r})$ with correlation length $\sigma_{\mathrm{sp}}= \SI{296(11:7)}{\nano\meter}$~\cite{hesse2025probingdisorderdriventopologicalphase}, comparable to the lattice spacing $a$. Its statistics follow the intensity distribution $p_{\overline{I}}=e^{-I/\overline{I}}/\overline{I}$ and are fully parametrized by the mean speckle intensity $\overline{I}$~\cite{Goodman_speckle}, hence by a mean disorder strength $\overline{V}_D$. A repulsive potential step projected by a digital micromirror device (DMD, $\lambda_{\mathrm{DMD}}=\SI{638}{\nano \meter}$) realizes a topological interface supporting chiral edge transport~\cite{Braun_2024}. Finally, localized wave packets (we assume the real-space standard deviation of the Gaussian wavefunction to equal the harmonic oscillator length $l_\mathrm{HO}=\SI{0.45}{\micro \meter}$) are prepared by loading about 500 atoms from a nearly non-interacting ($a_s=6a_0$) BEC into a single optical tweezer ($\lambda_\mathrm{tw}=\SI{1064}{\nano \meter}$), which can be positioned anywhere in the lattice.
\\
Figure~\ref{fig:Fig1}a illustrates the dynamics in the combined potential landscape throughout a characteristic experimental sequence. Release from the tweezer occurs either in the interior or along the edge of the honeycomb lattice, selectively populating bulk Floquet states or, if present, topological edge states. In a regime of moderate disorder strengths, wave packets initialized at opposite ends of the system (in red and grey) propagate chirally along the edges. In contrast, disorder slows down or even halts the expansion of bulk states (in blue). Imaging bulk expansions in the presence of disorder gives access to quantitative localization metrics, including observables that approximate the inverse participation ratio (IPR)~\cite{Bell_1970, Thouless_1974, Wegner_1980, Schwartz_2007}. In practice, topological edge modes with opposite group velocities are populated by flipping the chirality of the modulation scheme rather than positioning the tweezer along terminations with different handedness. Thus, a chiral signal can be obtained by subtracting absorption images for opposite chiralities of the modulation scheme from each other~\cite{Braun_2024, hesse2025probingdisorderdriventopologicalphase}. Representative absorption images for such experimental protocols in the anomalous Floquet regime are shown in Fig.~\ref{fig:Fig1}b, both in the absence (top panel) and presence (bottom panel) of a repulsive potential step.

Optical lattices provide a natural testbed for quantum simulation of transport phenomena in solid-state systems. In particular, previous work explored the coherent real-space evolution of initially confined bosonic~\cite{Weitenberg_2011, Ronzheimer_2013, Preiss_2015, Xia_2015, Wei_2023} and fermionic~\cite{Schneider_2012, Scherg_2018} clouds in static lattices, in both the non-interacting and strongly correlated regimes. In the absence of disorder and interactions, for conventional as well as Floquet Bloch states, the expansion velocity of a state at quasimomentum $\boldsymbol{q}$ within the $\mu$-th (quasi)energy band with dispersion $E_\mu(\boldsymbol{q})$ is given by $\boldsymbol{v}_{\mu}(\boldsymbol{q})=\nabla_{\boldsymbol{q}}E_{\mu}(\boldsymbol{q})/\hbar$~\cite{Dalibard2013OpticalLattices, Leung2020}. Abrupt release into the modulated honeycomb lattice from an optical tweezer with a waist of a few lattice constants prepares a superposition of Floquet Bloch states weighted by their respective overlap with the initial momentum-space wave packet. This superposition undergoes time evolution according to a distribution of velocities obtained from the dispersion of occupied states across the Brillouin zone.
\\
The top row of Fig.~\ref{fig:Fig2} shows the atomic density distribution imaged after release from the optical tweezer, along a cut through the parameter space at fixed modulation amplitude $m=0.25$. All profiles retain the $C_6$ symmetry of the lattice and display interference fringes, evidencing coherent expansion, but their shape changes markedly along the frequency axis. At the two lowest frequencies, the distribution is dominated by a central peak, at $\omega/(2\pi)=9$\,kHz it takes the form of a six-pointed star, around $10.5$--$11$\,kHz it develops into a ring, and at the two highest frequencies it becomes a uniformly filled hexagon. Along the cut, the system crosses three topological regimes, distinguished by the winding numbers $(W_0, W_\pi)$ of the two quasienergy gaps, the 0-gap $g^{(0)}=\min_{\boldsymbol q}\{g^{(0)}_{\boldsymbol q}\}$ inside the Floquet Brillouin zone and the $\pi$-gap $g^{(\pi)}=\min_{\boldsymbol q}\{g^{(\pi)}_{\boldsymbol q}\}$ at its boundary. The winding number of a gap counts the chiral edge modes traversing it, and the Chern number of a band is the difference between the winding numbers of the gaps below and above it~\cite{Rudner2013}. At $\omega/(2\pi)=5$\,kHz (orange shading) the system is in a Haldane-like (HL) regime with $(W_0,W_\pi)=(0,1)$ and Chern numbers $\mathcal{C}^\pm=\mp1$; from $7$ to $10.5$\,kHz (blue shading) it is in the anomalous Floquet (AF) regime with $(W_0,W_\pi)=(1,1)$, where both gaps host edge modes although $\mathcal{C}^\pm=0$; and at the three highest frequencies (green shading) it is in the Haldane (H) regime with $(W_0,W_\pi)=(1,0)$ and $\mathcal{C}^\pm=\pm1$.
\begin{figure*}[t!]
\centering
\includegraphics{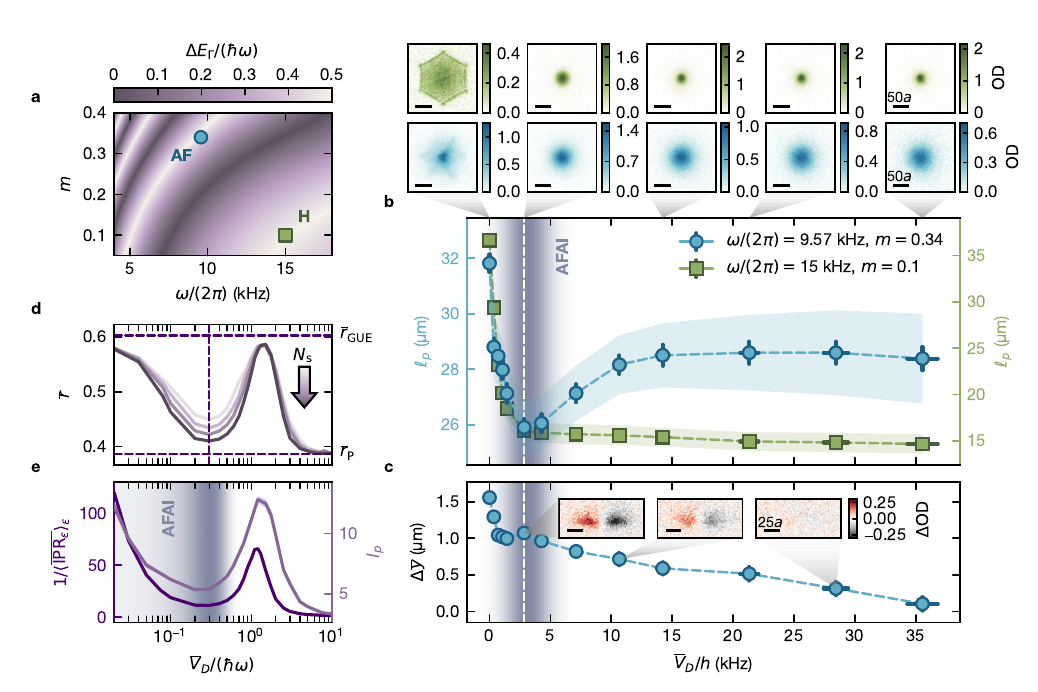}
\caption{\textbf{Signatures of an anomalous Floquet-Anderson insulating phase.} \textbf{a} Minimal energy gap $\Delta E_\Gamma= \mathrm{min}\{g^{(0)}_\Gamma,g^{(\pi)}_\Gamma\}$ at quasimomentum $\boldsymbol{q}=\Gamma$ across the phase diagram. The round blue marker at $(\SI{9.57}{\kilo \hertz},0.34)$ and the square green marker at $(\SI{15}{\kilo \hertz},0.1)$ indicate the modulation parameters used in \textbf{b} and \textbf{c}. \textbf{b} Participation length $\ell_p$ extracted from the average of 35--36 absorption images taken at $t_\mathrm{exp}=100T$ in the anomalous Floquet (blue) and Haldane (green) regimes. The white dashed line marks $\overline{V}_D/h \approx \SI{\fpeval{round(\specklecal * 0.2, 2)}}{\kilo\hertz} \approx 0.3\,\omega/(2\pi)$. Statistical error bars are obtained by bootstrapping; transparent bands display the systematic error (see SM). Insets: averaged expansion images at $\overline{V}_D/h \in \{0, \fpeval{round(\specklecal * 0.2, 2)}, \fpeval{round(\specklecal * 1, 2)}, \fpeval{round(\specklecal * 1.5, 2)}, \fpeval{round(\specklecal * 2.5, 2)}\}$~kHz. \textbf{c} Center-of-mass distance $\Delta \overline{y}$ quantifying chiral edge transport at $(\SI{9.57}{\kilo \hertz},0.34)$ as a function of disorder strength for $t_\mathrm{exp}=100T$. The differential signal is averaged over 115 disorder realizations. Insets: averaged edge state signal at $\overline{V}_D/h \approx \SI{\fpeval{round(\specklecal * 0.2, 2)}}{\kilo \hertz}$, $\SI{\fpeval{round(\specklecal * 0.75, 2)}}{\kilo \hertz}$ and $\SI{\fpeval{round(\specklecal * 2, 2)}}{\kilo\hertz}$. Error bars are obtained by bootstrapping. \textbf{d} Disorder-averaged level spacing ratio $\overline{r}$ of the two-band model with speckle disorder at $\hbar\omega/(hJ_0)=8.7$, for system sizes $N_{\mathrm{s}} \in \{336, 540, 836, 1560\}$ (increasing with arrow). Horizontal dashed lines mark the Poissonian ($\overline{r}\approx0.39$) and GUE ($\overline{r}\approx0.6$) limits; the vertical shading marks the local minimum around $\overline{V}_D=0.3\,\hbar\omega$. \textbf{e} Inverse of the quasienergy- and disorder-averaged IPR, $1/\langle \overline{\mathrm{IPR}_\varepsilon}\rangle_\varepsilon$, of the Floquet eigenstates (left axis) and dimensionless participation length $l_p$ after $n=50$ cycles of numerical time propagation of a Gaussian wave packet in the same model (right axis, see SM). $l_p^2$ counts the effective number of lattice sites occupied by the wave packet. Numerical data in \textbf{d} and \textbf{e} are averaged over 100 disorder realizations. The shaded region indicates the standard error of the mean of $l_p$. The minimum of $\ell_p$ in the anomalous Floquet regime (\textbf{b}) coincides with the disorder strength at which the model predicts localization of the bulk states (\textbf{d}, \textbf{e}) and finite chiral edge transport (\textbf{c}), consistent with an AFAI phase.
}
\label{fig:Fig3}
\end{figure*}
\\
To connect these shapes to the Floquet band structure, we compute the quasienergy bands of the driven lattice from a six-band Floquet calculation and extract a pair of $s$-like bands adiabatically connected to the static $s$-bands (see SM), onto which we project the initial wave packet, modeled as a Gaussian in momentum space. The simulated profiles in the bottom row of Fig.~\ref{fig:Fig2} are then obtained by propagating each Bloch state ballistically with its group velocity $\boldsymbol v_\mu(\boldsymbol q)\, t_\mathrm{exp}$. This construction neglects interference between different Bloch states, so it reproduces the envelope of the measured distributions but not the fringes, which are a genuine coherence effect. We find excellent agreement: the characteristic shapes follow directly from the dispersion. On either side of the HL--AF transition, a moat-like dispersion of the lowest band~\cite{Bracamontes2022, Sedrakyan2014, Sedrakyan2015} produces a large weight of states with vanishing group velocity, hence the central peak. Close to the AF--H transition at $\omega/(2\pi)=10.8$\,kHz, the bands become nearly linear around the closing gap, so that most populated states share the same speed and expand in a ring-like shape. In the H regime, the dispersion approaches that of the static honeycomb lattice and the distribution becomes a filled hexagon. This understanding of the clean system paves the way for using wave packet expansion as a probe of bulk localization once disorder is present.

We now introduce speckle disorder and fix the modulation parameters deep inside the AF and H regimes (see markers in Fig.~\ref{fig:Fig3}a), where the minimal energy gap $\Delta E_\Gamma$ at quasimomentum $\Gamma$ approaches its maximum value of $0.5\,\hbar\omega$. This places both measurements far from any gap closing, since disorder shifts the topological phase transitions~\cite{hesse2025probingdisorderdriventopologicalphase}, and compares the two regimes at equal gap size in units of $\hbar\omega$. To quantify bulk localization experimentally, we evaluate the inverse participation area~\cite{Schwartz_2007}
\begin{equation}
\mathcal{I}=\frac{\sum_{i,j}\mathrm{OD}_{ij}^2}{\left(\sum_{i,j}\mathrm{OD}_{ij}\right)^2},
\label{eq:IPA}
\end{equation}
where $\mathrm{OD}_{ij}$ denotes the integrated optical density binned into camera pixels $(i,j)$. The corresponding participation length $\ell_p=1/\sqrt{\mathcal{I}}$ measures the effective width of the cloud, which approximately follows the inverse IPR of the bulk Floquet states, as discussed below.
\\
In the AF regime, we observe a non-monotonic dependence of $\ell_p$ on the disorder strength (Fig.~\ref{fig:Fig3}b): it first decreases, attains a local minimum around $\overline{V}_D/h=\SI{2.84}{\kilo\hertz}\approx0.3\,\omega/(2\pi)$ and increases again towards a plateau at larger disorder strengths. This characteristic profile is independent of the expansion time for the timescales probed in the experiment (see SM). In the H regime, in contrast, $\ell_p$ decreases monotonically towards its value in the strong-disorder limit, as it does in the static honeycomb lattice (see SM), consistent with disorder driving the system directly into a trivial Anderson insulator.
\\
The defining feature of an AFAI is the coexistence of bulk localization with chiral transport along a topological interface, which breaks down only at larger $\overline{V}_D$~\cite{hesse2025probingdisorderdriventopologicalphase}. We therefore quantify chiral edge transport at the same modulation parameters $(\SI{9.57}{\kilo \hertz},0.34)$ as speckle disorder is introduced, by measuring the differential center-of-mass displacement $\Delta\overline{y}$ of the two drive chiralities, defined as in Ref.~\cite{hesse2025probingdisorderdriventopologicalphase}. Indeed, we observe a clear chiral edge signal at the dip of the $\ell_p$ curve, $\overline{V}_D/h\approx0.3\,\omega/(2\pi)$, where bulk expansion is most strongly suppressed (Fig.~\ref{fig:Fig3}c). For stronger disorder, the chiral signal decreases continuously and topological transport eventually breaks down. In the H regime, $\Delta\overline{y}$ instead decays together with $\ell_p$ (see SM), as expected for a direct transition into a topologically trivial regime.
\\
To relate these observations to the behavior expected for an AFAI phase, we determine numerically whether the modulation scheme supports an AFAI and at which disorder strength bulk localization is expected. To this end, we extend a two-band tight-binding model previously studied with uniformly distributed, spatially uncorrelated disorder~\cite{Dutta_2024} to a speckle potential with finite spatial correlation, and evaluate localization metrics as a function of its strength. A standard observable, connecting Anderson localization to random matrix theory and quantum chaos, is the level spacing ratio (LSR) $\overline{r}$, the mean ratio of consecutive quasienergy spacings~\cite{Titum_2016, Dutta_2024, Guhr_1998, Oganesyan_2007, Atas_2013}. In the thermodynamic limit it approaches the Gaussian unitary ensemble (GUE) value $\overline{r}_{\mathrm{GUE}}=2\sqrt{3}/\pi-1/2\approx0.6$ for delocalized and the Poissonian value $\overline{r}_{\mathrm{P}}=2\ln2-1\approx0.39$ for localized states. In the case of uniformly distributed disorder, the modulation scheme used here was found to support an AFAI at intermediate disorder strengths, separated from the trivial Anderson insulator at strong disorder by a critical point~\cite{Dutta_2024}. This localization--delocalization--localization sequence appears in $\overline{r}$ as a dip towards $\overline{r}_{\mathrm{P}}$, a peak towards $\overline{r}_{\mathrm{GUE}}$ at the critical point, and a final decay to $\overline{r}_{\mathrm{P}}$. Figure~\ref{fig:Fig3}d shows that a numerical speckle pattern yields the same sequence: a local minimum around $\overline{V}_D/(\hbar\omega)\approx0.3$, critical delocalization at $\overline{V}_D/(\hbar\omega)\approx1.2$, and a subsequent decrease to $\overline{r}_{\mathrm{P}}$, which sets in at larger $\overline{V}_D$ than for uniform disorder. The local minimum approaches $\overline{r}_{\mathrm{P}}$ for increasing number of lattice sites $N_{\mathrm{s}}$ and extrapolates to it in the thermodynamic limit (see SM), consistent with complete localization of the bulk states~\cite{Titum_2016, Suntajs_2023}.
\\
An alternative, real-space perspective on the localization of the Floquet eigenstates is provided by the IPR~\cite{Bell_1970, Thouless_1974, Wegner_1980, Schwartz_2007}. For a Floquet eigenstate $|\psi_\varepsilon\rangle$ with quasienergy $\varepsilon$ on a finite system of $N_{\mathrm{s}}$ sites, $\mathrm{IPR}_{\varepsilon}=\sum_{i=1}^{N_{\mathrm{s}}}|\langle \mathbf{r}_i|\psi_\varepsilon\rangle|^4$ approaches unity for a wavefunction localized on a single lattice site and $1/N_{\mathrm{s}}$ in the opposite limit of complete delocalization. The inverse of its quasienergy- and disorder-averaged value, $1/\langle\overline{\mathrm{IPR}_\varepsilon}\rangle_\varepsilon$, follows the trend of $\overline{r}$, exhibiting a dip around $\overline{V}_D/(\hbar\omega)=0.3$ (Fig.~\ref{fig:Fig3}e). Neither the LSR nor the IPR is within experimental reach, since the experiment probes the Floquet eigenstates only through the expansion of an initially localized wave packet $|\varphi_0\rangle$. Therefore, we simulate stroboscopic propagation of $|\varphi_0\rangle$, $|\varphi(nT)\rangle=\hat{U}(T)^n|\varphi_0\rangle$ with the Floquet propagator $\hat{U}(T)$, and evaluate the dimensionless participation length $l_p=(\sum_{i=1}^{N_{\mathrm{s}}}|\langle \mathbf{r}_i|\varphi(nT)\rangle|^4)^{-1/2}$ after $n=50$ cycles. As shown in Fig.~\ref{fig:Fig3}e, $l_p$ reproduces the disorder dependence of $1/\langle\overline{\mathrm{IPR}_\varepsilon}\rangle_\varepsilon$, which establishes the participation length of expansion profiles as a probe of the localization of the bulk Floquet states. The measured minimum of $\ell_p$ thus coincides with the disorder strength at which the two-band model predicts the bulk states to be localized, and the chiral edge transport observed there provides the second ingredient of an AFAI. The subsequent increase of $\ell_p$ follows the predicted approach towards the critical point. The expected re-localization at strong disorder, however, is not resolved: instead, $\ell_p$ saturates, which we attribute to disorder-induced hybridization with the $p$-bands, setting in only at large $\overline{V}_D$.

\begin{figure}[t!]
\centering
\includegraphics[width=\columnwidth]{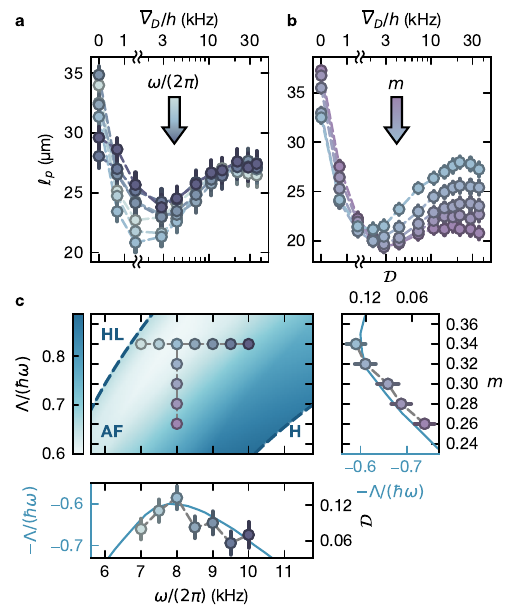}
\caption{\textbf{AFAI dip contrast across modulation parameters.}
\textbf{a} Participation length $\ell_p$ as a function of disorder strength for modulation frequencies $\omega/(2\pi) \in \{7, 7.5, 8, 8.5, 9, 9.5, 10\}\,\si{\kilo\hertz}$ at fixed $m=0.34$. \textbf{b} Same as \textbf{a} for modulation amplitudes $m \in \{0.26, 0.28, 0.3, 0.32, 0.34\}$ at fixed $\omega/(2\pi)=\SI{8}{\kilo\hertz}$. In both panels the color gradient follows the arrow and $\ell_p$ is extracted from disorder-averaged absorption images after $t_\mathrm{exp}=75T$ (30 and 45 averages per point in \textbf{a} and \textbf{b}, respectively). For clarity, the disorder axis is subdivided into a linear segment at weak disorder strengths up to $\overline{V}_D/h=\SI{1.5}{\kilo\hertz}$ and a logarithmic segment extending to the strong-disorder limit, separated by the axis break. Statistical error bars are obtained by bootstrapping. \textbf{c} $s$-pair bandwidth $\Lambda$ (see text) across the phase diagram. The dashed lines mark the cuts of \textbf{a} (horizontal) and \textbf{b} (vertical); markers indicate the measured points. Right and bottom subplots: dip contrast $\mathcal{D}$, Eq.~(2), extracted from the curves in \textbf{a} and \textbf{b} (right axis), together with $-\Lambda$ along the same cuts (left axis). Within the anomalous Floquet regime, larger dip contrast coincides with smaller $s$-pair bandwidth.
}
\label{fig:Fig4}
\end{figure}

Having identified the dip signature in Fig.~\ref{fig:Fig3}, we now investigate its robustness as the drive parameters are varied across the AF regime. The localization properties of a bulk band depend sensitively on its dispersion: flatter bands reduce the kinetic energy scale relative to disorder, favoring localization~\cite{Nagaoka_1985}. Across the AF regime, the bandwidth varies substantially with the modulation parameters, which limits the region of parameter space that can host an AFAI~\cite{Dutta_2024}. Figure~\ref{fig:Fig4}a displays $\ell_p$ as a function of $\overline{V}_D$ for varying $\omega/(2\pi)$. The local minimum persists for all modulation frequencies and is most pronounced at $\omega/(2\pi)=\SI{8}{\kilo\hertz}$, slightly away from the maximum of $\Delta E_\Gamma$ (cf.\ Fig.~\ref{fig:Fig3}a) and from the condition $\hbar\omega/(hJ_0)=8.7$ considered in Ref.~\cite{Dutta_2024}. Its position along the disorder axis, $\overline{V}_D/h \approx \SI{\fpeval{round(\specklecal * 0.15, 2)}}{\kilo\hertz}\approx0.27\,\omega/(2\pi)$ at $\SI{8}{\kilo\hertz}$, depends only weakly on the modulation frequency. All curves converge in the strong-disorder limit, consistent with the expansion time being fixed in units of the modulation period. Fixing $\omega/(2\pi)=\SI{8}{\kilo\hertz}$, we next vary the modulation amplitude (Fig.~\ref{fig:Fig4}b). The minimum becomes shallower with decreasing $m$ but persists down to $m=0.26$.

To quantify these observations, we introduce the dip contrast
\begin{equation}
\mathcal{D}=\frac{\ell_\mathrm{plateau}-\ell_\mathrm{min}}{\ell_\mathrm{plateau}+\ell_\mathrm{min}},
\label{eq:dipcontrast}
\end{equation}
where $\ell_\mathrm{plateau}$ denotes the mean participation length for $\overline{V}_D/h>\SI{15}{\kilo\hertz}$ and $\ell_\mathrm{min}$ the minimum of the curve. As a measure of the joint bandwidth of the $s$-band pair, we define $\Lambda=\Lambda_1+\Lambda_2+\min\{g^{(0)},g^{(\pi)}\}$, where $\Lambda_{1,2}$ are the widths of the two $s$-like Floquet bands obtained from the six-band calculation (see SM). The motivation is that at the disorder strengths of the dip, $\overline{V}_D$ is comparable to or exceeds the smaller of the two gaps, such that disorder effectively merges the two $s$-bands into a single band of width $\Lambda$, separated from its Floquet copies by the larger gap $\hbar\omega-\Lambda$. Figure~\ref{fig:Fig4}c displays $\Lambda$ across the AF regime together with $\mathcal{D}$ along the two cuts of Fig.~\ref{fig:Fig4}a,b. Along both cuts, we find that $\mathcal{D}$ is anti-correlated with $\Lambda$, i.e., the narrower the $s$-band pair, the more pronounced the dip. Band flatness alone, however, does not produce the signature. At the Haldane point of Fig.~\ref{fig:Fig3}, the $s$-band pair is of comparable width, $\Lambda/(\hbar\omega)\approx0.5$, yet $\ell_p$ decreases monotonically.

In conclusion, we have observed a non-monotonic dependence of bulk expansion on disorder strength in the anomalous Floquet regime of a tunneling-modulated honeycomb lattice. The minimum coincides with the disorder strength at which a two-band model with speckle disorder predicts localization of all bulk Floquet states. This signature coexists with chiral transport along a topological interface and is absent in the Haldane regime of the same lattice. It persists across the anomalous Floquet regime, with a contrast that is largest where the $s$-band pair is narrowest. Together, these observations support the existence of an AFAI phase in this system. They were made possible by uniting, in a single apparatus, Floquet engineering of the band structure, optical disorder with tunable strength, and real-space imaging of wave packets released from a controllable tweezer both in the bulk and at a programmable topological interface. Resolving the predicted re-localization into the trivial Anderson insulator at strong disorder and probing the quantized pumping that defines the AFAI~\cite{Titum_2016, Kundu2020} are natural next steps in the single-particle regime. The same toolbox also gives access to the dynamics of Dirac-like wave packets, which are predicted to evade Anderson localization altogether in smooth disorder~\cite{Bardarson2007, Nomura2007} and to display Klein tunneling through disorder barriers~\cite{Zhu2009, Kim2019}.
\\
The AFAI is further a crucial starting point for realizing interacting anomalous Floquet topological phases of matter, which have no static counterpart. MBL can persist under periodic driving at frequencies above the local many-body energy scales~\cite{Ponte2015, Lazarides2015, Abanin2016, Bordia2017} and offers a realistic route to protect the anomalous topology against heating, giving rise to chiral Floquet phases of MBL bosons~\cite{Po2016, Fidkowski_2019} and to interacting versions of the AFAI~\cite{Nathan2019}, insofar as MBL is stable in two dimensions, which is still under debate~\cite{Choi2016, DeRoeck2017}. %
Outside the MBL regime, heating in driven many-body systems~\cite{Dalessio2014, Lazarides2014, Mallayya2019, Boulier2019, Naeger2020} remains one of the key limitations. Several strategies to overcome this challenge have been demonstrated or proposed, including counterdiabatic driving~\cite{Claeys2019, Schindler2024}, two-tone drives with a weak auxiliary tone at a multiple of the main frequency that suppresses multi-photon excitations~\cite{Viebahn2021, Sandholzer2022, Chen2025}, and multipolar driving sequences with algebraically long prethermal lifetimes~\cite{Zhao2021}. The anomalous Floquet bands studied here are moreover of interest for interacting systems in their own right: the moat-like dispersion of the lowest band~\cite{Bracamontes2022} is predicted to suppress Bose condensation and to favor chiral spin liquids and composite-fermion states related to fractional quantum Hall physics~\cite{Sedrakyan2014, Sedrakyan2015}. %
Strongly correlated topological states have recently been realized with ultracold atoms in Floquet-engineered lattices, including few-particle fractional quantum Hall states~\cite{Leonard2023, Kwan2026} and strongly interacting Meissner phases in large bosonic flux ladders~\cite{Impertro2025}. This demonstrates that strongly correlated Floquet phases are within experimental reach, making the interacting AFAI a realistic target for future studies.

\bigskip
\begin{acknowledgments}
We thank W.~Hofstetter and his team, E.~Berg, R.~Moessner, M.~Bukov and G.~Roati for insightful discussions. This work was funded by the Deutsche Forschungsgemeinschaft (DFG, German Research Foundation) via Research Unit FOR2414 under project number 277974659 and Research Unit FOR5688 under project number 521530974. The work was further supported under Germany’s Excellence Strategy – EXC-2111–390814868, by the Alfried Krupp von Bohlen und Halbach Foundation, and under the Horizon Europe programme HORIZON-CL4-2022-QUANTUM-02-SGA via the project 101113690 (PASQuanS2.1).
\end{acknowledgments}
\bigskip

\bibliography{bibliography}

\cleardoublepage

\section*{Supplemental Material}
\setcounter{figure}{0}
\setcounter{equation}{0}
\setcounter{section}{0}
\setcounter{table}{0}

\renewcommand{\thefigure}{S\arabic{figure}}
\renewcommand{\theHfigure}{S\arabic{figure}} %
\renewcommand{\theequation}{S.\arabic{equation}}
\renewcommand{\thesection}{S.\Roman{section}}
\renewcommand{\thetable}{S\arabic{table}}

\begin{itemize}[labelwidth=3em, labelsep=0.5em, leftmargin=!, align=left]
\contentsitem{sec:exp_sequence}{Experimental sequence}
\contentsitem{sec:image_processing}{Image processing routine}
\contentsitem{sec:dis_cal}{Disorder strength calibration}
\contentsitem{sec:six_band}{Numerical methods: bulk expansion from Floquet group velocities}
\contentsitem{sec:numerics_LSR}{Numerical methods: level spacing ratio}
\contentsitem{sec:numerics_IPR}{Numerical methods: inverse participation ratio}
\contentsitem{sec:numerics_propagation}{Numerical methods: bulk expansion by Floquet propagation}
\contentsitem{sec:loc_static}{Extended data: localization dynamics in static honeycomb lattice}
\contentsitem{sec:ext_data_Haldane}{Extended data: edge state signal in Haldane regime}
\contentsitem{sec:ext_data_timeseries}{Extended data: participation length vs.\ expansion time}
\end{itemize}

\FloatBarrier

\section{Experimental sequence}
\label{sec:exp_sequence}

All experimental results discussed in the main text are obtained by performing the sequence described in Ref.~\cite{hesse2025probingdisorderdriventopologicalphase}. After loading a $^{39}$K Bose-Einstein condensate into the crossed optical dipole trap (ODT), the optical tweezer potential is ramped up adiabatically to a final loading value over about $\SI{30}{\milli \second}$, during which the optical dipole trap is simultaneously ramped down to expel the majority of the condensate not loaded in the tweezer trap. While doing so, the scattering length is ramped up to $a_{s,\mathrm{load}}= 13 a_0$ to maximize the number of atoms loaded in the optical tweezer. Subsequently, the tweezer power is reduced to its final value, resulting in the radial trapping frequency $\omega_r=2\pi \times \SI{1.3}{\kilo \hertz}$, while the ODT is ramped back up to its previous depth. In this trapping configuration, the lattice and disorder potentials are ramped up adiabatically within $\SI{20}{\milli \second}$, as well as the repulsive potential step to a barrier height of $V_\mathrm{DMD}/h = \SI{13.2}{\kilo \hertz}$ in case the sequence aims at imaging edge dynamics. In this joint potential landscape, the scattering length is ramped down to the value of interest for expansion dynamics, $a_{s,\mathrm{exp}} = 6a_0$, $\SI{4}{\milli \second}$ before the optical tweezer is abruptly switched off. Finally, absorption imaging is performed in situ after an expansion time $t_{\mathrm{exp}}$, with all potentials being left on except for the disorder potential, turned off about $\SI{10}{\micro \second}$ before imaging, given that the corresponding green light leaks onto the EMCCD camera and therefore affects the absorption imaging process. 

\section{Image processing routine}
\label{sec:image_processing}

\begin{figure}[t!]
\centering
\includegraphics{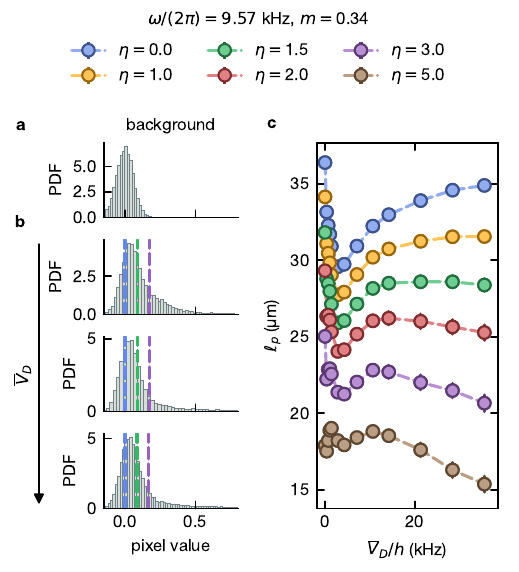}
\caption{\textbf{Dependence of the participation length on the noise-reduction threshold.}
\textbf{a} Probability density for pixel values in a background region of interest (ROI) containing no atomic signal.
\textbf{b} Pixel value distributions for disorder-averaged, background-subtracted absorption images in the selected ROI for increasing disorder strengths $\overline{V}_D/h=\SI{0}{\kilo\hertz}$, $\overline{V}_D/h=\SI{\fpeval{round(\specklecal * 0.2, 2)}}{\kilo\hertz}$, $\overline{V}_D/h=\SI{\fpeval{round(\specklecal * 1, 2)}}{\kilo\hertz}$, at a modulation frequency $\omega/(2\pi)= \SI{9.57}{\kilo \hertz}$ and a modulation amplitude $m=0.34$, corresponding to the dataset shown in Fig.~3b of the main text.
The noise reduction routine discards all pixel values that fall below $\eta\cdot\sigma_\mathrm{BG}$.
Vertical lines illustrate the thresholds for $\eta=0$ (blue), $\eta=1.5$ (green) and $\eta=3$ (purple).
\textbf{c} Disorder dependence of the participation length $\ell_p$ extracted using thresholds $\eta\in\{0,1,1.5,2,3,5\}$. Although the threshold affects the absolute participation length, the local minimum near $\overline{V}_D/h=\SI{\fpeval{round(\specklecal * 0.2, 2)}}{\kilo\hertz}$ persists over a range of moderate thresholds.}
\label{fig:lp_vs_eta}
\end{figure}

For each combination of modulation parameters, disorder strength, and expansion time $(\omega,m,\overline{V}_D,t_{\mathrm{exp}})$, we extract the participation length from the experimental images using the following procedure: First, we average the experimental images over different disorder realizations to improve the signal-to-noise ratio, which is otherwise insufficient for a reliable extraction of the participation length at the expansion times of interest. Independent disorder realizations are generated by rotating a diffuser illuminated off-axis by the speckle beam.
We restrict the analysis to a square region of interest (ROI) roughly centered around the tweezer position, whose size is chosen such that it includes the entirety of the most expanded density distribution among the dataset of interest.
A square frame of twenty pixels width around the ROI is used as a region without atoms to extract (i) the mean background offset and (ii) the standard deviation of the background-subtracted image $\sigma_{\mathrm{BG}}$, which is used in the next step. Prior to computing the inverse-participation area of the disorder-averaged, background-subtracted image, we take an additional step to minimize the impact of noise on the result: all pixels falling below a threshold of $\eta\cdot\sigma_{\mathrm{BG}}$ are set to zero, such that they do not contribute to the evaluation, as these pixels' signal-to-noise ratio is not large enough.
Finally, $\mathcal{I}$ is obtained from the processed image as described in the main text.
\\
The influence of the noise reduction threshold $\eta$, applied in the last step of the evaluation routine, is illustrated in Fig.~\ref{fig:lp_vs_eta} using the same dataset analyzed in Fig.~3b of the main text, for modulation parameters ($\SI{9.57}{\kilo \hertz}$, 0.34).
In particular, Fig.~\ref{fig:lp_vs_eta}a shows the probability density function (PDF) of pixel values within the background ROI in order to estimate the noise floor.
Similarly, the disorder-averaged optical density in the square signal ROI after background subtraction yields the PDFs shown in Fig.~\ref{fig:lp_vs_eta}b for three disorder strengths, increasing from top to bottom.
At every disorder strength, the distributions for noise and signal overlap in a considerable range of pixel values, preventing a straightforward choice of $\eta$. Different $\eta\cdot \sigma_\mathrm{BG}$ are highlighted as vertical dashed lines. Figure~\ref{fig:lp_vs_eta}c shows the corresponding participation length $\ell_p=1/\sqrt{\mathcal{I}}$ for increasing disorder strength $\overline{V}_D$.
For thresholds below approximately $\eta=2$, which avoid clipping most of the signal pixels associated with the atomic distribution, all curves exhibit a local minimum at the same value of $\overline{V}_D/h$, followed by an increase at larger disorder strengths, differing only in the absolute scale of $\ell_p$. At the largest disorder strengths, these curves do not decrease again but saturate. As discussed in the main text, we attribute this plateau to disorder-induced hybridization with the $p$-bands: the hybridized states inherit the much larger tunneling of the $p$-bands, so that their localization length greatly exceeds that of the $s$-band states and they expand into a dilute halo around the localized core. The optical density of this halo falls below large thresholds, which is why $\ell_p$ decreases again at large $\overline{V}_D$ for $\eta>2$.
\\
The noise reduction threshold chosen for the evaluation of all datasets presented in the main text is $\eta=1.5$. To obtain a systematic error, we vary $\eta \in \{1.25,1.5,1.75 \}$ and define the systematic error as half the difference between the maximum and the minimum value resulting from the three evaluations. The systematic error is shown in Fig.~3b as the shading surrounding the data points.

\section{Disorder strength calibration}
\label{sec:dis_cal}

We convert the voltage setpoints used to stabilize the speckle laser intensity into the disorder energy scale $\overline{V}_D/h$, expressed in kHz, using the calibration procedure described in Ref.~\cite{hesse2025probingdisorderdriventopologicalphase}. This procedure relies on Kapitza–Dirac diffraction of the BEC by a pulsed speckle potential. We compare the measured decay rates of the zeroth diffraction order with those obtained from numerical solutions of the Gross–Pitaevskii equation (GPE).
\\
Over the entire period during which the measurements presented in this work were taken, we recorded seven independent speckle calibration datasets with slightly varying imaging detunings and atom numbers, in order to sample systematic variations introduced by these settings. The calibration constants obtained by fitting the measured decay rates to GPE simulations are $\alpha_i\pm\Delta\alpha_i\in\{16.16\pm0.87,\,14.92\pm0.62,\,12.60\pm0.43,\,12.94\pm0.48,\,12.89\pm0.41,\,14.83\pm0.39,\,15.19\pm0.61\}$~kHz/V, where $\Delta\alpha_i$ denotes the fit uncertainty. All disorder axes are calibrated with their mean, $\bar{\alpha}\approx\SI{14.2}{\kilo\hertz\per\volt}$. The scatter of the individual calibrations about this mean, $\sigma_\alpha=\sqrt{\sum_i(\alpha_i-\bar{\alpha})^2/(N-1)}\approx\SI{1.4}{\kilo\hertz\per\volt}$ with $N=7$, exceeds the typical fit uncertainty of a single calibration, $\sqrt{\sum_i\Delta\alpha_i^2/N}\approx\SI{0.6}{\kilo\hertz\per\volt}$. The scatter therefore reflects run-to-run variations beyond the fit uncertainties, which we attribute to the different imaging conditions. The deliberate variation of these conditions across the calibration runs is what allows us to quantify this systematic effect: it is contained in the scatter $\sigma_\alpha$, and the resulting uncertainty of the mean, $\Delta\bar{\alpha}=\sigma_\alpha/\sqrt{N}\approx\SI{0.5}{\kilo\hertz\per\volt}$, includes it together with the statistical fit uncertainties, corresponding to a common uncertainty of about $4\%$ on all disorder axes.

\section{Numerical methods: bulk expansion from Floquet group velocities}
\label{sec:six_band}

\begin{figure*}[t]
\centering
\includegraphics{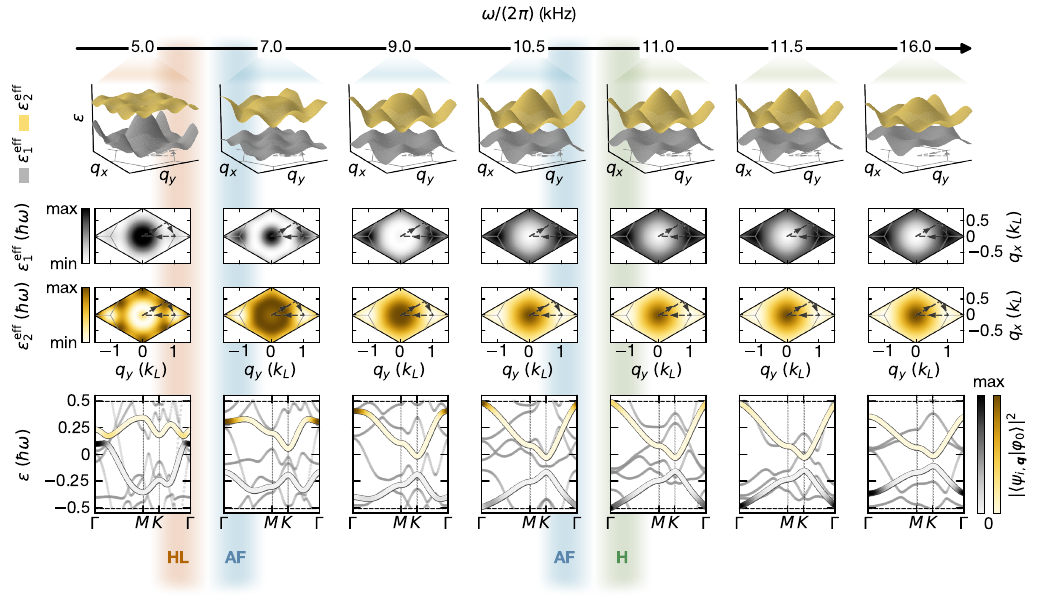}
\caption{\textbf{Effective $s$-like Floquet bands and initial state overlaps.} Top row: dispersions of the effective $s$-like Floquet bands $\varepsilon_{1}^{\mathrm{eff}}(\boldsymbol{q})$ (gray) and $\varepsilon_{2}^{\mathrm{eff}}(\boldsymbol{q})$ (yellow), identified using the projection algorithm across two-dimensional momentum space. The modulation frequency is fixed for every column and increases from left to right: $\omega/(2\pi) \in \{5,7,9,10.5,11,11.5,16 \}$~kHz. Second and third rows: projected dispersion of first and second effective band, respectively, shown on a rhomboid Brillouin zone. Bottom row: quasienergy dispersions along the high-symmetry path $\Gamma$-$M$-$K$-$\Gamma$, illustrated in the Brillouin zone (BZ) above. Light-gray background curves show all six eigenenergies obtained from the truncated Floquet Hamiltonian. For the two selected bands, the color represents the squared overlap $|\langle \psi_i(\boldsymbol{q}) | \varphi_0 \rangle|^2$ of the corresponding state with a real-space Gaussian wave packet with $\sigma_r=\SI{446}{\nano \meter}$ for the amplitude, modeling the initial state $|\varphi_0\rangle$ prior to release of the atoms from the optical tweezer.}
\label{fig:SM_bandproj}
\end{figure*}

This section provides detailed information about the numerical calculation of the expansion profiles displayed in the bottom row of Fig.~2. 
As in Ref.~\cite{Wintersperger_2020}, we project the Floquet spectra resulting from a six-band Floquet calculation onto two effective $s$-like bands, which are adiabatically connected to the $s$-bands of the static honeycomb lattice.
\\
Given the wavevectors $\boldsymbol{k}_0=(0,1)$, $\boldsymbol{k}_1=(-\sqrt{3}/2,-1/2)$, $\boldsymbol{k}_2=(\sqrt{3}/2,-1/2)$ of the three running waves that interfere to form the honeycomb lattice, the corresponding reciprocal lattice is spanned by $\boldsymbol{b}_1=\boldsymbol{k}_0-\boldsymbol{k}_1$ and $\boldsymbol{b}_2=\boldsymbol{k}_2-\boldsymbol{k}_0$, such that reciprocal lattice vectors read $\boldsymbol{G}=l_1\boldsymbol{b}_1+l_2\boldsymbol{b}_2$, with $l_1,l_2\in \mathbb{Z}$. The time-periodic Hamiltonian describing a Floquet drive with modulation frequency $\omega/(2\pi)$ and amplitude $m$ is given by introducing a time-dependent perturbation to the Hamiltonian for the static lattice $H_{\textrm{stat}}(\boldsymbol{q})$: 
\begin{equation}
    H^{(m,\omega)}(\boldsymbol{q},t) = H_{\textrm{stat}}(\boldsymbol{q}) + H_{\textrm{drive}}^{(m,\omega)}(\boldsymbol{q},t),
\end{equation}
with $H_{\textrm{drive}}^{(m,\omega)}(\boldsymbol{q},t) = H_{\textrm{drive}}^{(m,\omega)}(\boldsymbol{q},t+T)$. Both terms are expressed in a basis $\{|\boldsymbol{q}+\boldsymbol{G}\rangle\}$ of 225 plane waves, with $|l_i|\leq l_\mathrm{max}=7$. The time period $[0,T)$ is discretized into $N_\mathrm{steps}=100$ steps with time interval $\Delta t=T/N_\mathrm{steps}$: $t_j=j T/N_\mathrm{steps}$ for $j=0,...,N_\mathrm{steps}-1$. Subsequently, we diagonalize $H^{(m,\omega)}(\boldsymbol{q},t_j)$ at every $t_j$ and truncate the Hilbert space to the six lowest eigenstates $|\phi^{(m,\omega)}_n (\boldsymbol{q},t_j)\rangle$ with eigenenergies $E^{(m,\omega)}_n(\boldsymbol{q},t_j)$, $n=1,...,6$. The projected Hamiltonians in the six-band basis are labeled as $H_p^{(m,\omega)}(\boldsymbol{q},t_j)$. The Floquet propagator over one drive period at every quasimomentum is then finally constructed as the time-ordered product
\begin{equation}
    U_p^{(m,\omega)}(\boldsymbol{q}, T) = \prod_j e^{-\frac{i}{\hbar} H_p^{(m,\omega)}(\boldsymbol{q},t_j) \Delta t}.
\end{equation}
The six eigenvalues $\lambda_n^{(m,\omega)}(\boldsymbol{q})$ of $U_p^{(m,\omega)}(\boldsymbol{q}, T)$ contain the quasi-energies $\varepsilon_n^{(m,\omega)}(\boldsymbol{q}) \in \left( -\hbar\omega/2, \hbar\omega/2 \right]$ as eigenphases: $\lambda_n^{(m,\omega)}(\boldsymbol{q})=\exp\{-2\pi i \varepsilon_n^{(m,\omega)}(\boldsymbol{q})/(\hbar \omega)\}$.
\\
In order to select a pair of eigenvalues to form bands which are adiabatically connected to the two $s$-bands of the static honeycomb lattice, the six-band propagator is again projected to a two-band propagator $U_s^{(m,\omega)}(\boldsymbol{q}, T)$:
\begin{equation}
    U_s^{(m,\omega)}(\boldsymbol{q}, T) = P^{\dagger}U_p^{(m,\omega)}(\boldsymbol{q}, T)P,
\end{equation}
with the projector onto the $2\times 2$ subspace spanned by the two lowest bands of the Hamiltonian at the start of the modulation, $\hat{P}= |\phi_1^{(m,\omega)} (\boldsymbol{q}, 0) \rangle \langle \phi_1^{(m,\omega)} (\boldsymbol{q}, 0)| + |\phi_2^{(m,\omega)} (\boldsymbol{q}, 0) \rangle \langle \phi_2^{(m,\omega)} (\boldsymbol{q}, 0)|$. $U_s$ is no longer necessarily unitary, nor are its eigenvectors orthogonal, due to leakage into the subspace of the four higher $p$-bands. Solving the eigenproblem $U_s^{(m,\omega)}(\boldsymbol{q}, T) \psi_k^{(m,\omega)}(\boldsymbol{q}) =\lambda_k^{(m,\omega)}(\boldsymbol{q}) \psi_k^{(m,\omega)}(\boldsymbol{q}) $ yields two effective $s$-like Floquet bands $\varepsilon_{1,2}^{\mathrm{eff}}(\boldsymbol{q})$ and Floquet Bloch states, to be assigned to either the first or the second band. The assignment is fixed at $\Gamma$ for each modulation frequency from the known phase diagram and its band inversions, and propagated outwards to neighboring quasimomenta. Figure~\ref{fig:SM_bandproj} shows the dispersion of the two resulting effective bands across two-dimensional quasimomentum space and along the high-symmetry line $\Gamma$-$M$-$K$-$\Gamma$ for all modulation frequencies of interest. In the lowest row, the grey points represent the eigenphases of the six-band propagator, whereas the two colored bands consist of the eigenphases of the two-band propagator. Where $p$-subspace leakage is negligible, the latter fall on top of the former.
\\
We consider the overlap of the two effective bands with an initial Gaussian wavefunction $|\varphi_0 \rangle$ in order to identify which Floquet eigenstates are realistically populated in an experimental sequence upon abruptly releasing the tightly confined atomic wave packet from the optical tweezer into the modulated honeycomb lattice.
Given a real-space width $\sigma_r$, $\varphi_0 (\boldsymbol{r}) \propto e^{-r^2/(2\sigma_r^2)}$, the momentum-space wave packet follows as $\varphi_0(\boldsymbol{k}) \propto e^{-k^2\sigma_r^2/2}$. The Floquet states obtained above can be expressed as 
\begin{equation}
    | \psi_i^{(m,\omega)}(\boldsymbol{q}) \rangle = \sum_{\alpha=1,2} \left( \psi^{(m, \omega)}_i (\boldsymbol{q}) \right)_\alpha |\phi_\alpha^{(m, \omega)}(\boldsymbol{q},0)\rangle,
\end{equation}
where $\left( \psi^{(m, \omega)}_i (\boldsymbol{q}) \right)_\alpha$ defines the $\alpha$-th ($\alpha=1,2$) component of the eigenvector $\psi^{(m, \omega)}_i (\boldsymbol{q})$. The overlap weights are then obtained as
\begin{align}
    W_i^{(m,\omega)}(\boldsymbol{q}) &= |\langle\psi_i^{(m,\omega)}(\boldsymbol{q})|\varphi_0\rangle|^2.
\end{align}
Finally, to account for the varying loading phase, i.e., release time from the tweezer relative to time zero, we enforce $C_3$ quasimomentum symmetry on the weights obtained above by averaging over a set of rotations $\mathcal{R}$:
\begin{equation}
    \overline{W}^{(m,\omega)}_i(\boldsymbol{q})= \frac{1}{3} \sum_{R\in\mathcal{R}}W_i^{(m,\omega)}(R\boldsymbol{q}), \: \mathcal{R}=\{R_0,R_{2\pi/3},R_{4\pi/3}\}.
\end{equation}
The weights above quantify the contribution of each Floquet state to the initial wave packet and are represented by the color scale of the corresponding quasienergies in the lower row of Fig.~\ref{fig:SM_bandproj}. We use these to construct a weighted histogram of group velocities, from which we derive the numerical expansion profiles shown in Fig.~2 of the main text.
\\
\begin{figure}[t!]
    \centering
    \includegraphics{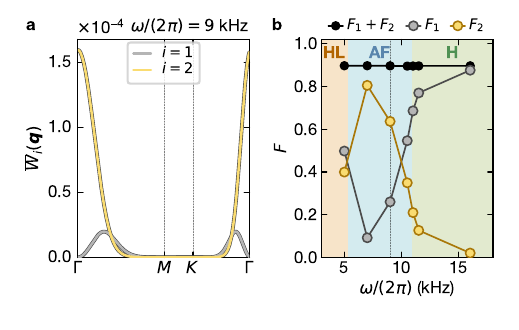}
    \caption{\textbf{Overlaps of effective Floquet eigenstate pair with initial state.} \textbf{a} Momentum-resolved initial-state weights for lower (grey) and upper (yellow) effective Floquet eigenstates along the corresponding bands. The finite momentum-space width causes the weights to vanish towards the Brillouin zone edges. \textbf{b} Fraction of overall initial state's six-band weight contained in the lower ($F_1$, grey) and upper effective band ($F_2$, yellow) resulting from the projection algorithm. The deviation of the sum $F_1+F_2$ from unity provides an estimate for leakage out of the effective $2\times 2$ subspace.}
    \label{fig:SM_band_proj_quant}
\end{figure}
A more quantitative analysis of the initial state projection is presented in Fig.~\ref{fig:SM_band_proj_quant}. In particular, Fig.~\ref{fig:SM_band_proj_quant}a displays overlap weights $\overline{W}^{(m,\omega)}_i(\boldsymbol{q})$ at a selected modulation frequency $\omega/(2\pi)= \SI{9}{\kilo \hertz}$. For varying modulation frequency across all three regimes of interest, Fig.~\ref{fig:SM_band_proj_quant}b displays the fraction of the initial state's six-band weight lying in the effective $s$-like Floquet band $i$:
\begin{equation}
    F_i^{(m,\omega)} = \frac{\sum_{\boldsymbol{q}\in \mathrm{BZ}}\overline{W}^{(m,\omega)}_i(\boldsymbol{q})}{\sum_{\boldsymbol{q}\in \mathrm{BZ}} \sum_{n=1}^6 |\langle \phi_n^{(m,\omega)}(\boldsymbol{q})|\varphi_0\rangle|^2}.
\end{equation}
In the expression above, the denominator is independent of the modulation frequency and quantifies to which extent the tweezer state is captured by the truncation to six bands. $1- (F_1+F_2)$ provides a measure of leakage into the $p$-subspace.

\begin{figure*}[t]
\centering
\includegraphics{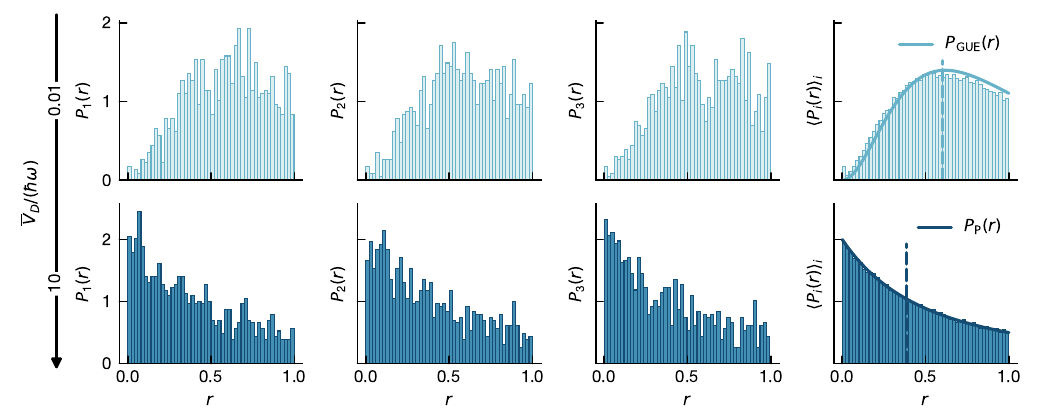}
\caption{\textbf{LSR distributions in the GUE and Poissonian limits.} The three leftmost columns show the probability distributions $P_i(r)$ for three individual disorder realizations on finite honeycomb lattices consisting of $N_\mathrm{s}=1144$ sites. The rightmost column shows the disorder-averaged probability distribution $\langle P_i(r) \rangle_i=\left(\sum_{i=1}^{100} P_i(r)\right)/100$. Upper row: at weak disorder, $\overline{V}_D/(\hbar\omega)=0.01$, the averaged distribution closely follows the Gaussian unitary ensemble (GUE) prediction (solid line), with mean $\overline{r}_\mathrm{GUE}\approx0.60$ (vertical dashed line). Lower row: at strong disorder, $\overline{V}_D/(\hbar\omega)=10$, the averaged distribution closely follows the prediction for Poissonian level statistics (solid line), with mean $\overline{r}_\mathrm{P}\approx0.386$ (vertical dashed line).}
\label{fig:SM_LSRs}
\end{figure*}

\section{Numerical methods: level spacing ratio}
\label{sec:numerics_LSR}

The following section summarizes the numerical approach to calculate the disorder-dependent LSR for the results shown in Fig.~3d in the main text. Disorder breaks the discrete translational symmetry of the honeycomb lattice, such that the ab-initio construction of a Hamiltonian in quasimomentum space (cf. Sec.~\ref{sec:six_band}) is no longer possible. Therefore, a real-space description of the system is required for numerical studies to evaluate localization metrics in a disordered honeycomb lattice. To this end, we employ the two-band tight-binding model introduced in Ref.~\cite{Dutta_2024}, which is qualitatively equivalent to the sinusoidal modulation of the lattice beam intensities realized experimentally. Specifically, in a tight-binding picture, the time-dependent tunneling rates along the three main directions of the honeycomb lattice are parametrized as
\begin{equation}
    J_i(t)=J_0 e^{F \cos{(\omega t + 2\pi/3(i-1))}}, \: i\in \{1,2,3\},
\end{equation}
where $J_0$ denotes the bare tunneling rate and $F$ parametrizes the modulation amplitude $m$ and therefore the bandwidth of the two bands.
Trotterizing the corresponding time-periodic Hamiltonian $\hat{H}(t)$ at $t_j=jT/N$ for $j=0,1,...,N-1$, the time-propagator over one drive period is obtained by
\begin{equation}
    \hat{U}(T)=\prod_{j=0}^{N-1}e^{-\frac{i}{\hbar}\left( \hat{H}(t_j)+\hat{V}_D\right)\frac{T}{N}},
\end{equation}
where $\hat{V}_D=\sum_{i=1}^{N_\mathrm{s}} V_i |i\rangle \langle i|$ is a spatial disorder operator diagonal in the lattice site basis with energy offsets $V_i$ sampled from a speckle pattern, as described in Ref.~\cite{hesse2025probingdisorderdriventopologicalphase}. For a specific disorder realization, the corresponding effective Hamiltonian follows as
\begin{equation}
    \hat{H}_{\mathrm{eff}}=\frac{i\hbar}{T} \ln{\left( \hat{U}(T) \right)}.
\end{equation}
Solving the eigenproblem $\hat{H}_\mathrm{eff}|\psi_n \rangle = \varepsilon_n |\psi_n \rangle $ yields the eigenbasis $\{ | \psi_n\rangle, \: n=1,2,...,N_{\mathrm{s}} \}$ and the Floquet eigenspectrum $\{ \varepsilon_n, \: n=1,2,...,N_{\mathrm{s}} \}$.

In random matrix theory, the level spacing ratio (LSR) provides a quantitative measure of the correlation between eigenvalue spacings of a random matrix. The same concept applies to disordered tight-binding Hamiltonians: given the eigenvalue spacings $\delta_n=\varepsilon_{n+1}-\varepsilon_n$, the corresponding ratios between consecutive spacings read
\begin{equation}
    r_n=\frac{\min\{\delta_{n+1},\delta_n\}}{\max\{\delta_{n+1},\delta_n\}}=\min\left\{\frac{\delta_{n+1}}{\delta_{n}},\frac{\delta_{n}}{\delta_{n+1}}\right\}.
\end{equation}
Sampling over the entire energy spectrum for various disorder realizations yields a probability distribution for the level-spacing ratio $P(r)$, characterized by its mean $\overline{r}$. Upon averaging over enough disorder realizations, $P(r)$ converges to a Gaussian unitary ensemble (GUE) distribution $P_\mathrm{GUE}(r)=81 \sqrt{3}/(2\pi)(r+r^2)^2/(1+r+r^2)^4$ in the limit of delocalized states. On the other hand, in the opposite limit of complete localization, $P(r)$ has been shown to converge to a Poissonian distribution $P_\mathrm{P}(r)=2/(1+r)^2$ \cite{Atas_2013, Titum_2016}. The corresponding statistical means are given by $\overline{r}_\mathrm{GUE}=2\sqrt{3}/\pi-1/2\approx0.6$ and $\overline{r}_\mathrm{P}=2\ln2-1\approx 0.39$, respectively.

Figure~\ref{fig:SM_LSRs} displays three exemplary probability distributions for the consecutive spacings ratio (three leftmost panels) as well as the corresponding disorder-averaged distributions obtained by considering 100 such realizations (rightmost panel), for the AFAI configuration $\hbar\omega/(hJ_0)=8.7$, in both limits of almost vanishing disorder ($\overline{V}_D/(\hbar\omega)=0.01$, top row) and very strong disorder ($\overline{V}_D/(\hbar\omega)=10$, bottom row). By varying the disorder strength between the extreme cases considered above, curves like those shown in Fig.~3d of the main text are obtained for a fixed system size.

\begin{figure}[t!]
    \centering
    \includegraphics{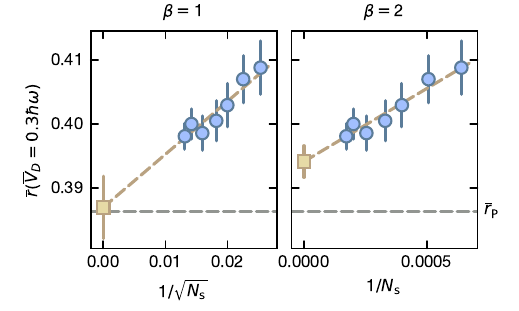}
    \caption{\textbf{Finite-size extrapolation of the level spacing ratio in the AFAI regime.} Blue points show the mean level spacing ratio $\overline{r}$ at fixed disorder strength $\overline{V}_D/(\hbar\omega)=0.3$ for increasing system sizes $N_\mathrm{s} \in \{1560, 1972, 2508, 3024, 3936, 4968, 5800\}$. The error bars represent the statistical error over 150 independent numerical disorder realizations. To estimate the LSR mean in the thermodynamic limit, we fit $\overline{r}=\overline{r}_{\infty}+ \gamma N_\mathrm{s}^{-\beta/2}$, assuming a linear dependence on inverse linear system size ($\beta=1$, left) or inverse system area ($\beta=2$, right). The yellow fits yield intercepts of $\overline{r}_{\infty}=0.3869\pm0.0024$ and $0.3941\pm0.0012$, respectively. Both values approach the Poissonian limit as highlighted by the horizontal dashed line $\overline{r}_\mathrm{P} \approx 0.386$. }
    \label{fig:SM_LSR_extrp}
\end{figure}

Figure~3d of the main text displays the disorder dependence of the LSR for increasing system size $N_\mathrm{s}$: $\overline{r}$ depends most sensitively on $N_\mathrm{s}$ around the local minimum for $\overline{V}_D/(\hbar \omega)=0.3$, in the disorder range expected to support the existence of an AFAI regime. A common approach to estimate the value of the LSR in the thermodynamic limit, $\overline{r}_{N_\mathrm{s}\rightarrow \infty}$, is to fit a power-law correction to the size-dependence in finite systems \cite{Binder_1997}:
\begin{equation}
    \overline{r}(L) \approx  \overline{r}_{N_\mathrm{s}\rightarrow \infty} + \gamma L^{-\beta}, 
\end{equation}
where $L$ stands for the linear system size, hence for $\sqrt{N_\mathrm{s}}$, $\gamma$ is a fit parameter and $\beta$ a model parameter. Figure~\ref{fig:SM_LSR_extrp} shows the LSR values obtained by averaging over 150 disorder realizations on a finite honeycomb lattice for an increasing number of sites $N_\mathrm{s}$ plotted against $L^{-1}=1/\sqrt{N_\mathrm{s}}$ ($\beta=1$, left panel) and $L^{-2}=1/N_\mathrm{s}$ ($\beta=2$, right panel). In both cases, the intercept $\overline{r}_{N_\mathrm{s}\rightarrow \infty}$ fit to the data appears close to the Poissonian limit, consistent with complete localization in the thermodynamic limit.

\begin{figure}[t!]
    \centering
    \includegraphics{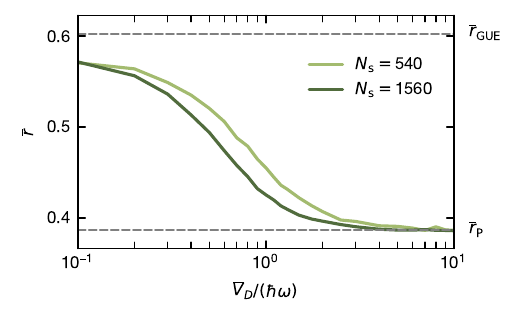}
    \caption{\textbf{Disorder-dependent level spacing ratio in the Haldane regime.} Deep in the Haldane regime ($\hbar \omega/(hJ_0)=20$), the mean level-spacing ratio $\overline{r}$, averaged over 100 disorder realizations, decreases monotonically with increasing disorder strength, in contrast to the behavior at $\hbar\omega/(hJ_0)=8.7$. It approaches the Poissonian limit at strong disorder, consistent with Anderson localization, and shows only weak dependence on the system size $N_\mathrm{s}$.}
    \label{fig:SM_Haldane_LSR}
\end{figure}

Finally, we evaluate the disorder dependence of the average LSR for varying system size deep in the Haldane regime, at $\hbar\omega/(hJ_0)=20$, for comparison with the anomalous regime at $\hbar\omega/(hJ_0)=8.7$ studied numerically in the main text. Figure~\ref{fig:SM_Haldane_LSR} shows the resulting disorder dependence after averaging over 100 disorder realizations for two system sizes $N_\mathrm{s}$: regardless of the system size, $\overline{r}$ displays a steady decay to the Poissonian limit in the regime of strong disorder.

\section{Numerical methods: inverse participation ratio}
\label{sec:numerics_IPR}

\begin{figure}[t!]
\centering
\includegraphics{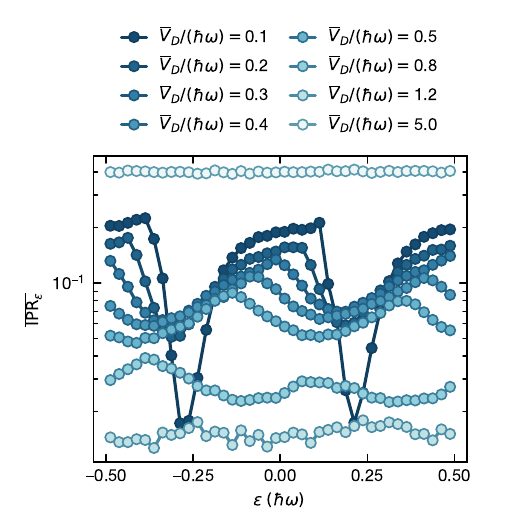}
\caption{\textbf{Disorder-dependent IPR across the Floquet Brillouin zone.} Quasienergy-resolved, disorder-averaged IPR $\overline{\mathrm{IPR}_\varepsilon}$ for progressively increasing disorder strength (shades of blue) in the anomalous regime ($\hbar \omega/(hJ_0)=8.7$). For disorder strengths insufficient to mix the two energy bands and bridge the gaps, the IPR retains a quasienergy-dependence. In contrast, an approximately constant value is approached upon mixing the bands, which is smaller for delocalized states (more lattice sites are occupied on average by eigenstates) and larger for localized states (fewer lattice sites are occupied on average by eigenstates). The dip signature highlighted in the main text manifests as the quasienergy average over the Floquet Brillouin zone, which displays a local minimum at the critical disorder strength $\overline{V}_D/(\hbar \omega)=1.2$ and subsequently increases towards the limit of strong disorder. Each data point is obtained by averaging over 100 disorder realizations.}
\label{fig:SM_IPR_vs_e}
\end{figure}

In this section, we illustrate the principles underlying the numerical calculation of the inverse participation ratio (IPR) for the data displayed on the left $y$-axis of Fig.~3e of the main text. The IPR provides a measure of the real-space spread of Floquet eigenstates across the Floquet Brillouin zone. For a given quasienergy $\varepsilon$, the corresponding IPR reads
\begin{equation}
    \mathrm{IPR}_\varepsilon = \sum_{\alpha=1}^{N_{\mathrm{s}}} \sum_{i=1}^{N_{\mathrm{s}}} |\langle i|\psi_\alpha\rangle|^4 \delta(\varepsilon - \varepsilon_\alpha),
\end{equation}
with the first sum running over eigenstates and the second over lattice sites. For a delocalized state at quasienergy $\varepsilon$, the corresponding disorder-averaged $\overline{\mathrm{IPR}_\varepsilon}$ is predicted to scale as the inverse of the system area $L^2$ in a two-dimensional system ($\overline{\mathrm{IPR}_\varepsilon} \sim L^{-2}$), whereas for an Anderson-localized state at quasienergy $\varepsilon$, $\overline{\mathrm{IPR}_\varepsilon}$ is independent of the system size~\cite{Dutta_2024}. For the modulation parameters of interest, namely $\hbar\omega/(hJ_0)=8.7$ and $F=2$ as for the numerical results presented in Fig.~3e of the main text, the quasienergy-dependence of the disorder-averaged IPR is illustrated in Fig.~\ref{fig:SM_IPR_vs_e} for selected disorder strengths.

The asymmetry of $\overline{\mathrm{IPR}}_\varepsilon$ about the quasienergy zero of the Brillouin zone, progressively more pronounced as $\overline{V}_D$ increases, stems from the skewness inherent to the exponential intensity distribution of optical speckles.
In fact, equivalent studies performed on the same two-band model in Ref.~\cite{Dutta_2024} for on-site disorder offsets $\{V_i\}_{i=1}^{N_\mathrm{s}}$ sampled from a uniform disorder distribution centered around zero energy resulted in curves symmetric about $\varepsilon/(\hbar \omega)=0$.
For speckle-like disorder, however, subtracting the mean on-site disorder $\langle V \rangle=\left(\sum_{i=^1}^{N_\mathrm{s}}V_i\right)/N_\mathrm{s}$ from the diagonal for each disorder realization does not suffice to restore the symmetry of the binned IPR about the Floquet Brillouin zone (FBZ) origin. Subtracting the mean leaves the skewness of the distribution unchanged: most sites have disorder offsets below the mean, while a small fraction have much larger offsets in the long positive tail.

\begin{figure}[t!]
\centering
\includegraphics{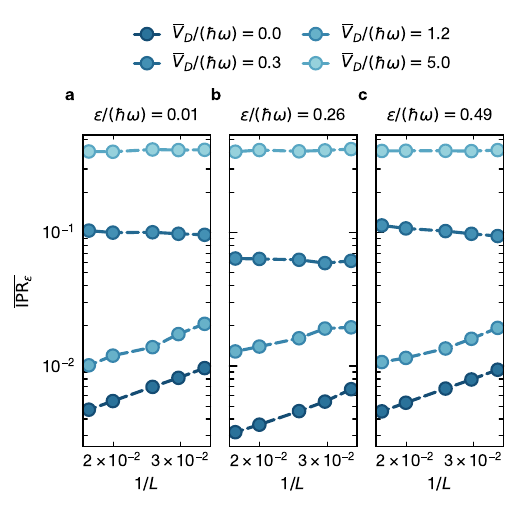}
\caption{\textbf{System-size dependence of the IPR in the anomalous Floquet regime.} Disorder-averaged inverse participation ratio $\overline{\mathrm{IPR}\varepsilon}$ as a function of inverse linear system size, $1/L= 1/\sqrt{N\mathrm{s}}$, at selected quasienergies: \textbf{a}, at the Floquet Brillouin zone (FBZ) center; \textbf{b}, halfway between its center and edge; and \textbf{c}, at its edge. At $\overline{V}_D/(\hbar\omega)=0.3$, the IPR shows no discernible size dependence at any of these quasienergies, consistent with bulk localization in the AFAI regime. In contrast, at the critical point separating the two localized regimes, $\overline{V}_D/(\hbar\omega)=1.2$, the IPR exhibits pronounced size dependence, consistent with delocalization. Each data point is obtained by averaging over 50 disorder realizations.}
\label{fig:SM_IPR_scaling}
\end{figure}

Figure~\ref{fig:SM_IPR_scaling} displays the size-dependence of $\overline{\mathrm{IPR}_\varepsilon}$ at the center of the FBZ ($\varepsilon/(\hbar\omega)=0$, left panel), halfway to the edge of the FBZ ($\varepsilon/(\hbar\omega)=0.25$, center panel), and at the edge of the FBZ ($\varepsilon/(\hbar\omega)=0.5$, right panel).
As expected, in the two localized regimes at intermediate $\overline{V}_D/(\hbar \omega)=0.3$ and strong disorder strength $\overline{V}_D/(\hbar \omega)=5$, the IPR is independent of system size, regardless of the quasienergy considered. In contrast, at the critical point separating these regimes, $\overline{V}_D/(\hbar\omega)=1.2$, $\mathrm{IPR}_{\varepsilon}$ exhibits a pronounced dependence on system size, consistent with delocalization of the corresponding eigenstates. This disorder strength also coincides with local maxima in the LSR and the inverse IPR.

\begin{figure*}
\centering
\includegraphics{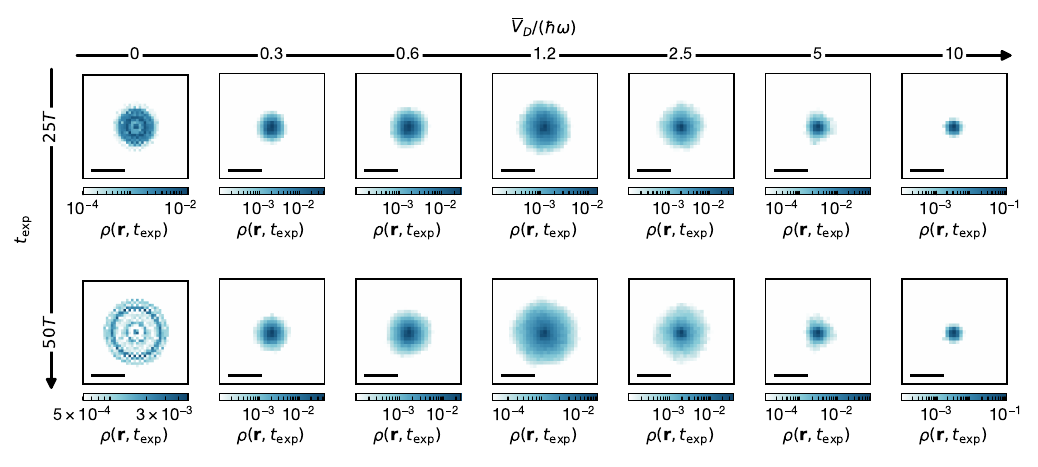}
\caption{\textbf{Simulated bulk expansions in the AFAI regime.} At $\hbar \omega/(hJ_0)=8.7$ we time-propagate a Gaussian wave packet centered on a lattice site with a standard deviation equal to the harmonic oscillator length $\sigma_r = l_{\mathrm{HO}} = \SI{446}{\nano \meter} \approx 1.55a$ for the wavefunction, using the Floquet operator $\hat{U}(T)$ and calculate the density distribution $\rho(\mathbf{r},t)$ at integer multiples of the modulation period. Varying the disorder strength $\overline{V}_D$, the dip structure in the spatial extent of the cloud becomes evident for long enough expansion times. The upper and lower rows display disorder-averaged (100 individual realizations per finite disorder strength) numerical expansions for an expansion time of 25 and 50 modulation cycles, respectively. Black scale bars in both rows indicate $20a\approx\SI{6}{\micro\meter}$.}
\label{fig:SM_num_props}

\vspace{5mm}

\centering
\includegraphics{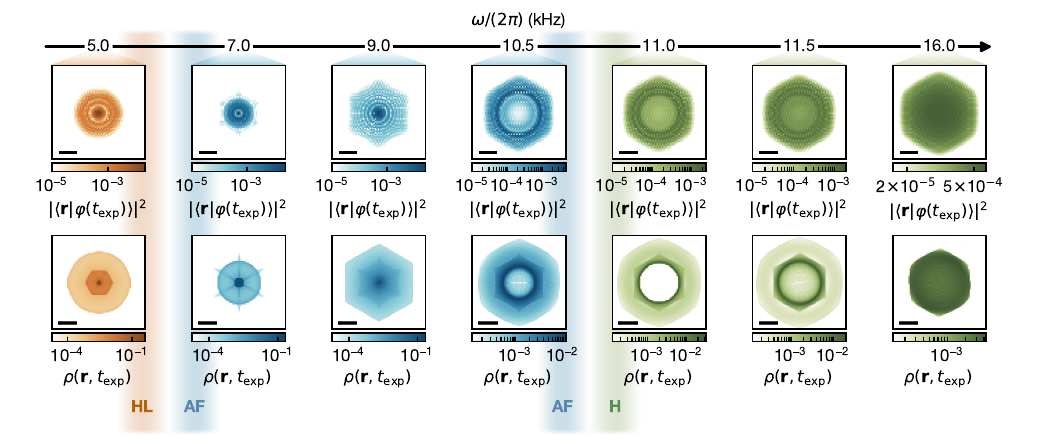}
\caption{\textbf{Comparison of numerical expansion profiles obtained using two approaches.}
Upper row: real-space propagation of an initial Gaussian wavefunction centered on a lattice site with a standard deviation $\sigma_r = l_{\mathrm{HO}} = \SI{446}{\nano \meter}\approx 1.55a$ on a finite lattice using the Floquet operator of a two-band tight-binding model.
Lower row: expansion profiles reconstructed from the group velocities of two effective Floquet bands obtained from a six-band momentum-space model, weighted by their overlaps with the initial wave packet, given the same real-space width as for the upper row. We use an expansion time of $t_\mathrm{exp}=\SI{2.5}{\milli\second}$ for both approaches, chosen to accommodate the finite lattice size in the real-space calculation. Black scale bars in both rows indicate $20a\approx\SI{6}{\micro\meter}$.}
\label{fig:SM_exps_comp}
\end{figure*}

\section{Numerical methods: bulk expansion by Floquet propagation}
\label{sec:numerics_propagation}

In Fig.~3e of the main text, we compare the theoretical participation ratio with the spatial extent of numerical density profiles $\rho(\mathbf{r}, t_\mathrm{exp})$ obtained by stroboscopic Floquet evolution up to $t_\mathrm{exp}=nT,~n\in \mathbb{N},$ in the same two-band model considered in the two previous sections.
To reproduce the expansion of tightly confined wave packets in the bulk of a modulated honeycomb lattice numerically, we consider finite lattice systems and evaluate the Floquet propagator $\hat{U}(T)$ by enforcing closed boundary conditions on the instantaneous tight-binding Hamiltonians $\hat{H}(t_j)$ throughout the procedure summarized above.
As extensively described in Ref.~\cite{hesse2025probingdisorderdriventopologicalphase}, this provides insight into the stroboscopic time evolution undergone by an initial numerical state $| \varphi_0\rangle$ localized to one or very few lattice sites in the bulk, away from the system boundaries.
After an evolution time of $n$ modulation periods, the initial state is mapped onto $|\varphi(nT)\rangle=\hat{U}(nT)| \varphi_0 \rangle= \hat{U}(T)^n | \varphi_0 \rangle$.
We extract a numerical equivalent of the participation length obtained from the averaged absorption images to track the dependence on the disorder strength $\overline{V}_D$ after a fixed number of Floquet cycles:
\begin{equation}
    l_p=\left(\sum_{i=1}^{N_{\mathrm{s}}}|\langle \mathbf{r}_i|\varphi(nT)\rangle|^4\right)^{-1/2}.
\end{equation}
As shown in Fig.~3e of the main text, the numerical $l_p$ reproduces the qualitative disorder dependence of the inverse of the disorder- and quasienergy-averaged IPR, supporting the use of its experimental counterpart $\ell_p$ as a probe of localization.
Figure~\ref{fig:SM_num_props} shows the disorder-averaged density distributions $\rho(\boldsymbol{r},nT)=\overline{|\langle\boldsymbol{r}|\varphi(nT)\rangle|^2}$ for the numerical expansions at evolution times of $25T$ (top row) and $50T$ (bottom row). The finite system sizes accessible numerically limit the evolution times we can study.
As visible at both expansion times, the radial extent of the spatial profile decreases down to its minimum around $\overline{V}_D/(\hbar \omega)=0.3$, increases again and peaks at $\overline{V}_D/(\hbar \omega)=1.2$, before decaying monotonically towards the strong-disorder limit. 

\begin{figure}[t]
\centering
\includegraphics{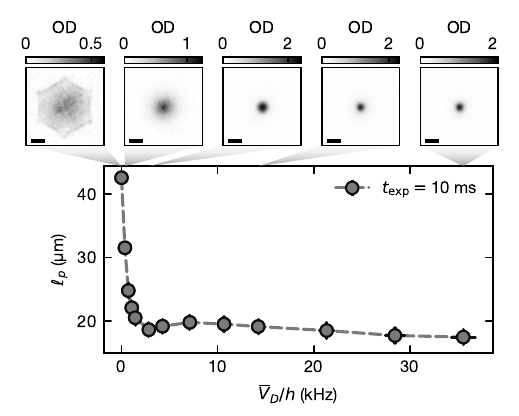}
\caption{\textbf{Bulk dynamics in disordered static honeycomb lattices.} Participation length of bulk expansions in the absence of a Floquet drive for increasing disorder strength and expansion time $t_\mathrm{exp}=\SI{10}{\ms}$. Similar to the Haldane regime considered in the main text, $\ell_p$ decays steadily towards its value in the strong-disorder limit. Top row: corresponding disorder-averaged in situ images of bulk expansions at disorder strengths of $\overline{V}_D/h \in \{ 0, \fpeval{round(\specklecal * 0.025, 2)}, \fpeval{round(\specklecal * 0.2, 2)}, \fpeval{round(\specklecal * 1, 2)}, \fpeval{round(\specklecal * 2.5, 2)} \}$~kHz. The black scale bars indicate distances of $50a$. Participation lengths are extracted from average absorption images obtained by averaging over 39 individual disorder realizations. Error bars are obtained by bootstrapping. The noise reduction threshold used for the analysis is $\eta=1.5$.}
\label{fig:SM_statlat}
\end{figure}

Real-space Floquet propagation in a two-band tight-binding model as introduced above
provides a fundamentally different method for numerically predicting the expansion dynamics in the bulk of modulated honeycomb lattices than the one summarized in Sec.~\ref{sec:six_band}.
To compare the two procedures at equivalent modulation parameters, we use a tight-binding model with the same temporal dependence of the tunneling amplitudes as above, but with a different parametrization. This parametrization has been shown in Ref.~\cite{hesse2025probingdisorderdriventopologicalphase} to describe the drive at $m=0.25$, matching the modulation amplitude used for the dataset of interest in Fig.~2 of the main text and its numerical reproduction.
Figure~\ref{fig:SM_exps_comp} presents a quantitative comparison of the resulting density distributions at a fixed expansion time of $t_\mathrm{exp}=\SI{2.5}{\milli \second}$: in the top row, $|\langle \mathbf{r}|\varphi(t_\mathrm{exp}) \rangle|^2$ is obtained by projection of the time-propagated state for the two-band model onto the site basis, whereas in the bottom row $\rho(\mathbf{r},t_\mathrm{exp})$ is obtained as extensively described in Sec.~\ref{sec:six_band} considering six bands.
Evidently, the impact of hybridization with $p$-bands on the dispersion of the effective $s$-bands obtained from band projection compared to a bare two-band model varies considerably along the modulation frequency axis.
For instance, in the case of the two lowest modulation frequencies considered, hybridization seems to enhance band dispersion and therefore result in larger group velocities at the relevant quasimomenta.
On the other hand, towards the high-frequency limit, the largest modulation frequency $\omega/(2\pi)=\SI{16}{\kilo \hertz}$ displays a slightly less spread expansion profile when taking into account higher-lying bands.

\begin{figure}[t!]
\centering
\includegraphics{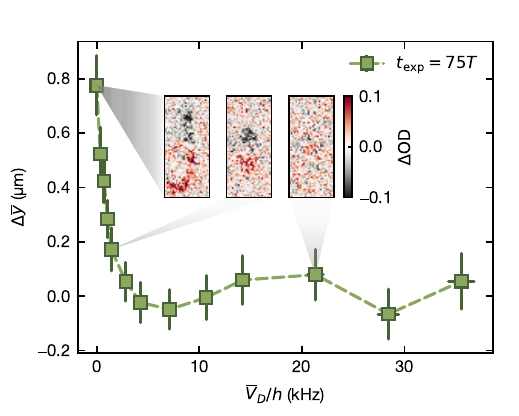}
\caption{\textbf{Disorder dependence of chiral edge transport in the Haldane regime.} We quantify chiral edge state transport through the differential center-of-mass signal $\Delta\overline{y}$, measured after an expansion time of $t_\mathrm{exp}=75T$ in the Haldane regime ($\omega/(2\pi)=\SI{15}{\kilo\hertz}$, $m=0.1$). Unlike in the anomalous regime, $\Delta\overline{y}$ decreases with increasing disorder in qualitative agreement with the participation length $\ell_p$, suggesting that disorder drives a direct transition into a topologically trivial regime. The insets show representative edge state signals without disorder (left, $\overline{V}_D/h=\SI{0}{\kilo\hertz}$), just before the breakdown of topological transport (center, $\overline{V}_D/h=\SI{\fpeval{round(\specklecal * 0.1, 2)}}{\kilo\hertz}$), and in the trivial regime (right, $\overline{V}_D/h=\SI{\fpeval{round(\specklecal * 1.5, 2)}}{\kilo\hertz}$). We average over 83 disorder realizations for each drive chirality. Statistical error bars on $\Delta\overline{y}$ are obtained by bootstrapping.}
\label{fig:SM_Haldane_es}
\end{figure}

\section{Extended data: localization dynamics in the static honeycomb lattice}
\label{sec:loc_static}

To provide a reference for the behavior of the participation length extracted from disordered bulk expansions in the modulated lattice, Fig.~\ref{fig:SM_statlat} displays the result of an equivalent set of measurements in the absence of the Floquet drive, i.e., in a static honeycomb lattice. In particular, $\ell_p$ is extracted for varying $\overline{V}_D$ after an expansion time of $t_\mathrm{exp}=\SI{10}{\ms}$, qualitatively reproducing the steady decay observed for the Haldane regime for $\omega/(2\pi)=\SI{15}{\kilo \hertz}$, $m=0.1$ (cf. Fig.~3 in the main text).

\section{Extended data: edge state signal in the Haldane regime}
\label{sec:ext_data_Haldane}

\begin{figure}[!t]
\centering
\includegraphics{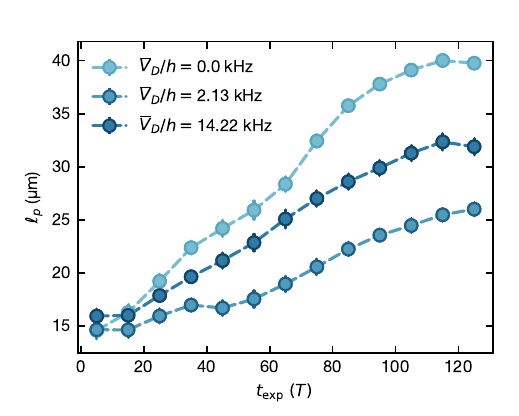}
\caption{\textbf{Time evolution at selected disorder strengths.} Participation length of disorder-averaged expansion profiles in the anomalous regime ($\omega/(2\pi)=\SI{8}{\kilo \hertz}$, $m=0.34$) for varying expansion time. Darker colors correspond to larger disorder strength. For every expansion time except the shortest, the atomic cloud expands more for $\overline{V}_D/h= \SI{\fpeval{round(\specklecal * 1, 2)}}{\kilo\hertz}$ than for $\overline{V}_D/h= \SI{\fpeval{round(\specklecal * 0.15, 2)}}{\kilo\hertz}$, in proximity to the dip signature along the disorder axis. Error bars are obtained by bootstrapping over the 39 individual disorder realizations per data point and are almost always smaller than the marker. The noise reduction threshold used for the analysis is $\eta=1.5$.}
\label{fig:SM_time_exp}
\end{figure}

Here, we investigate the decay of chiral edge-mode transport in the Haldane regime by analyzing the differential center-of-mass (COM) signal at $\omega/(2\pi)=\SI{15}{\kilo\hertz}$ and $m=0.1$. We present this analysis in the Supplemental Material because Ref.~\cite{hesse2025probingdisorderdriventopologicalphase} discusses the same observable in detail for different modulation parameters within the Haldane regime. As described in the aforementioned reference, the edge state signal in the Haldane regime is optimized by increasing the radial trapping frequency of the tweezer to $\omega_r=2\pi \times \SI{2}{\kilo \hertz}$ and lowering the barrier height to $V_\mathrm{DMD}/h=\SI{2.8}{\kilo \hertz}$.
\\
Figure~\ref{fig:SM_Haldane_es} displays the results of this analysis, equivalent to the one carried out in the main text for the edge state decay in the anomalous regime (cf. Fig.~3c of the main text), with an expansion time consisting of 75 modulation cycles.
As opposed to in the anomalous regime, the disappearance of the edge state signal with increasing disorder strength follows the bulk localization behavior probed by the expansions displayed in Fig.~3b of the main text. 
Despite abundant disorder averaging and imaging at a shorter expansion time in units of modulation periods than for equivalent datasets acquired in the anomalous regime presented in the main text, the edge state signal in the Haldane regime at $(\SI{15}{\kilo \hertz},0.1)$ is characterized by a significantly smaller signal-to-noise ratio. We primarily attribute this to a reduced overlap of edge states occurring at finite quasimomentum, in proximity to the Brillouin zone edge with the initial state released from the tweezer, centered around $\Gamma$ in momentum space, as well as to the small absolute size of the energy gap at $K$ hosting the edge mode.

\section{Extended data: participation length vs. expansion time}
\label{sec:ext_data_timeseries}
All measurements in the main text are taken at a fixed expansion time. The expansion time is limited by the harmonic confinement and by heating in the modulated lattice, which is most pronounced at large modulation amplitudes such as $m=0.34$.
Figure~\ref{fig:SM_time_exp} shows the time dependence of the participation length for bulk expansions at $\omega/(2\pi)=\SI{8}{\kilo\hertz}$ and $m=0.34$. We compare three disorder strengths: $\overline{V}_D/h=\SI{0}{\kilo\hertz}$, corresponding to the clean system; $\overline{V}_D/h=\SI{\fpeval{round(\specklecal * 0.15, 2)}}{\kilo\hertz}$, within the dip associated with the AFAI regime; and $\overline{V}_D/h=\SI{\fpeval{round(\specklecal * 1, 2)}}{\kilo\hertz}$, near the critical delocalization predicted numerically. At all measured expansion times except the shortest, $5T$, the atomic cloud has the largest spatial extent in the clean system and the smallest at intermediate disorder. The ordering of the three curves is therefore independent of the expansion time, consistent with the non-monotonic dependence of $\ell_p$ on disorder reported in the main text.

\end{document}